\documentclass[aps,twocolumn,pra,superscriptaddress,nofootinbib,floatfix]{revtex4-2}
\usepackage{amsmath,amssymb,bm}
\usepackage{graphicx}
\usepackage{epstopdf}
\usepackage{latexsym}
\usepackage[normalem]{ulem} 
\usepackage{appendix}
\usepackage{dsfont}
\usepackage[usenames,dvipsnames]{color}
\usepackage{hyperref}
\usepackage{natbib}
\usepackage[dvipsnames]{xcolor}
\usepackage{braket}
\usepackage{mathtools}
\usepackage{empheq}
\usepackage{mdframed}
\usepackage{tcolorbox}
\usepackage{environ}
\NewEnviron{equationS}{%
\begin{align}\begin{split}
  \BODY
\end{split}\end{align}
}
\begin{document}

\newcommand{\Rev}[1]{{\color{blue}{#1}\normalcolor}} 
\newcommand{\Com}[1]{{\color{red}{#1}\normalcolor}} 
\newcommand*\widefbox[1]{\fbox{\hspace{2em}#1\hspace{2em}}}

\newcommand{\ketbra}[2]{|#1\rangle\langle #2|}
\newcommand{\normord}[1]{\mathopen{:}\,#1\,\mathopen{:}}

\author{D. Barberena}
\affiliation{JILA, NIST, Department of Physics, University of Colorado,  Boulder, CO 80309, USA}
\affiliation{Center for Theory of Quantum Matter, University of Colorado, Boulder, CO 80309, USA}
\author{Ana Maria Rey}
\affiliation{JILA, NIST, Department of Physics, University of Colorado, Boulder, CO 80309, USA}
\affiliation{Center for Theory of Quantum Matter, University of Colorado, Boulder, CO 80309, USA}

\title{Critical steady states of all-to-all squeezed and driven superradiance: An analytic approach}
\date{\today}

\begin{abstract}
We analyze the properties across steady state phase transitions of two all-to-all driven-dissipative spin models that describe possible dynamics of $N$ two-level systems inside an optical cavity: squeezed superradiance and driven superradiance. We show that the finite size behaviour around the critical points can be captured correctly by carefully identifying the relevant non-linearities in the Holstein-Primakoff representation of spin operators in terms of bosonic variables. With these tools, we calculate analytically various observables across the phase transitions and obtain their finite size scalings, including numerical prefactors. In particular, we look at the amount of spin squeezing carried by the steady states, of relevance for quantum metrology applications, and describe in analytical detail the mechanism by which the optimal spin squeezing acquires logarithmic corrections that depend on the system size. We also demonstrate that the logarithmic nature of these corrections is difficult to characterize through numerical procedures for any experimentally realistic and/or simulable values of particle number. We complement all of our analytical arguments with numerical benchmarks.
\end{abstract}
\maketitle  
\section{Introduction}
Studying the behaviour of quantum systems in the presence of decoherence and dissipation~\cite{breuer2002theory} is of paramount importance in the field of quantum technologies, both for practical and fundamental reasons. On the practical side, mitigating the effects of unwanted sources of decoherence is a necessary requirement for applications in quantum metrology~\cite{Cappellaro2017}, simulation~\cite{Altman2021,Georgescu2014} and computation~\cite{Bharti2022}. On the fundamental side, combining dissipation with coherent processes and/or drives can lead to novel kinds of behaviour~\cite{Vega2017,Helmrich2020,Bohnet2012,Meiser2009,Gong2018}, both dynamically~\cite{Iemini2018,Zhu_2015,Owen2018,Link2020} and in steady state conditions~\cite{Eisert2010,Maghrebi2016,Young2020,Marino2016,LeBoite2013,Rota2019,Sieberer2013}, and these insights can then be used as tools for more pragmatic endeavors.

An important avenue of research in driven-dissipative systems is devoted to understanding the steady states towards which these systems relax at long times. In particular, these steady states can undergo phase transitions when parameters of the system are varied. The strong reorganization of quantum and classical fluctuations that occurs near these phase transition points can then be utilized for e.g. quantum metrology applications, and the possibility of accessing these resources by just waiting for a system to relax constitutes a very appealing prospect for the preparation of entangled states. Importantly, tuning the system close to a transition point can be done not only by controlling coherent processes but also by deliberately engineering dissipation sources~\cite{Kraus2008,Diehl2008,Verstraete2009}. 

A class of driven-dissipative systems where steady states are of practical relevance is provided by all-to-all spin models~\cite{Carmichael_1980,Morrison2008,Lee2014,Iemini2018,Kessler2012,Ferreira2019,Pavlov2023}, where the generation of steady state spin squeezing~\cite{Kitagawa1993,MA2011} (useful for e.g. accurate timekeeping~\cite{Robinson2022,PedrozoPenafiel2020,Schulte2020}) is a well-documented effect. The amount of spin squeezing attainable is typically diagnosed using a variety of controlled approximations (mean field theory, Holstein-Primakoff~\cite{Holstein1940}, etc.) which give the correct answer when $N$, the number of spins, goes to infinity. In practice, the optimal squeezing is estimated numerically because it occurs close to phase transition points, where these analytic approaches ordinarily fail~\cite{Hirsch2013}. The common expectation is that this optimal value shows finite-size scaling and thus improves with $N$ according to a power law dependence. This is also true of generic observables and quantum phase transitions in closed all-to-all systems~\cite{J.Reslen_2005,Bastarrachea-Magnani_2014,Nahmad-Achar_2013}, though in that case renormalization~\cite{Dusuel2004,J.Vidal_2006} and field theory~\cite{DallaTorre2013b} techniques have been used to get analytical control.

Building on techniques used for ground state transitions~\cite{Liberti2006,Liberti2010,Titum2020}, in this paper we show that observables of all-to-all systems close to steady-state phase transition points can be calculated analytically by using the Holstein-Primakoff approximation consistently. In particular, we focus on the optimal spin squeezing in two distinct all-to-all spin models that have favorable metrological properties~\cite{Pavlov2023,Groszkowski2022}. We find that there are non-power-law corrections in $N$ that arise due to the optimization process and that are unique to the open quantum system setting, where steady states can be mixed. These corrections behave logarithmically when $N$ is very large, but clean observation of this trend requires working with $N>10^{23}$ particles, which is outside the scope of any realistic numerical simulation and partly explains the discrepancies in reported power law exponents~\cite{Lee2014,Barberena2019,Somech2022,Groszkowski2022}.

Our work is organized as follows: in Sec.~\ref{sec:Models} we introduce the mathematical models that we will study and give a small overview of their common mathematical properties. In Sec.~\ref{sec:SDM} we analyse squeezed superradiance, where an ensemble of atoms interacts with squeezed vacuum light~\cite{DallaTorre2013,AGARWAL1989,Gutierrez2023}, and whose critical properties are known to behave differently depending on the parity of $N$~\cite{Groszkowski2022}. In line with this, we perform independent analyses for each of these cases. Finally, in Sec.~\ref{sec:CRF} we study driven superradiance, where an ensemble of atoms is subjected to the competition between an external laser drive and collective decay of excitations. This is a model that has also been studied in other contexts under the name of cooperative resonance fluorescence~\cite{Carmichael_1980,Drummondo1980,Walls1980}. 

\section{Models}\label{sec:Models}
We consider two different driven-dissipative models, described by the following master equations
\begin{align}
        \partial_t\hat{\rho}&=\frac{\Gamma'}{N}\mathcal{D}\big(\hat{S}_x-i \zeta\hat{S}_y\big)\hat{\rho}\label{eqn:DallaTorre}\\[3pt]
        \partial_t\hat{\rho}&=-i\big[\Omega\hat{S}_x,\hat{\rho}\big]+\frac{\Gamma}{N} \mathcal{D}(\hat{S}^-)\hat{\rho},\label{eqn:CollectiveResonanceFluoresence}
\end{align}
where $\hat{\rho}$ is the density matrix of the system, $\mathcal{D}(\hat{O})\hat{\rho}=\hat{O}\hat{\rho}\hat{O}^\dagger-\{\hat{O}^\dagger\hat{O},\hat{\rho}\}$ is the standard dissipation superoperator, $\hat{S}_{x,y,z}=\sum_{i=1}^N\hat{\sigma}_{x,y,z}^i/2$ are collective spin operators, $\hat{\sigma}_{x,y,z}^i$ are Pauli matrices describing the $i^{\text{th}}$ two-level system, $\hat{S}^-=\hat{S}_x-i\hat{S}_y$, and $\Omega$, $\Gamma$, $\Gamma'$ and $\zeta$ are system parameters that can be varied to access the different steady state phases present in these models. Steady states $\hat{\rho}_{\text{ss}}$ are defined as the solutions to Eqs.~(\ref{eqn:DallaTorre}) and~(\ref{eqn:CollectiveResonanceFluoresence}) that satisfy $\partial_t\hat{\rho}=0$. These models were chosen because of their relevance in the literature, their favorable metrological properties, and the possibility of implementing them experimentally using minimal ingredients. We include schematic depictions of specific implementations in cavity systems in Figs.~\ref{fig:SDMModel}(a) and~\ref{fig:CRFModel}(a), and we explain them in more detail in the relevant sections of the paper.
 
Both models conserve (in the strong sense~\cite{Buca_2012,lieu2020symmetry}) the spin length operator, $\hat{S}^2=\hat{S}_x^2+\hat{S}_y^2+\hat{S}_z^2$, which allows us to focus on a single symmetry sector of Hilbert space, characterized by $\hat{S}^2=(N/2)(N/2+1)$. The states satisfying this condition constitute the ``Dicke manifold", are symmetric under permutation of the atoms and span an $N+1$ dimensional representation of the $SU(2)$ algebra generated by $\hat{S}_{x,y,z}$. This reduction in the size of the relevant sector of Hilbert space means that numerical simulations for large values of $N\sim 10^4$ are possible, even in the presence of dissipation. Since we will be working exclusively in the Dicke manifold, we will make a small digression to point out important features about these states before addressing the specific properties of the models we will study.
\subsection{Properties of Dicke manifold}
A typical basis for the Dicke manifold is given by $\ket{m}$, which are eigenstates of $\hat{S}_z$ satisfying $\hat{S}_z\ket{m}=m\ket{m}$, with $|m|\leq N/2$. These states can also be represented graphically as a distribution on the surface of a collective Bloch sphere of length $N/2$ by means of various quasi-probability functions. In this paper we will exclusively use the Husimi distribution, defined as
\begin{equation}
Q_{\hat{\rho}}(\theta,\phi)=\frac{1}{4\pi}\braket{\theta,\phi|\hat{\rho}|\theta,\phi},    
\end{equation}
where $\theta$ and $\phi$ are zenith and azimuthal angles in spherical coordinates, and $\ket{\theta,\phi}$ are spin coherent states~\cite{JMRadcliffe_1971}:
\begin{equation}
    \ket{\theta,\phi}=(\cos\theta/2)^{N}e^{\tan(\theta/2) e^{i\phi}\hat{S}^-}\ket{m=N/2}.
\end{equation}
In particular, the Husimi function of the spin coherent state $\ket{\theta,\phi}$ is highly concentrated along the $\theta,\phi$ direction on the Bloch sphere and is rotationally symmetric around this direction. The distribution is of size $\sim\sqrt{N}$ transverse to the Bloch vector direction.  
\section{Model I: Squeezed Superradiance (SSR)}\label{sec:SDM}
The model given in Eq.~(\ref{eqn:DallaTorre}), which we reproduce here
\begin{equation}
    \partial_t\hat{\rho}=\frac{\Gamma'}{N}\mathcal{D}\big(\hat{S}_x-i \zeta\hat{S}_y\big)\hat{\rho},
\end{equation}
is named ``squeezed superradiance"~\cite{Koppenhofer2023,munoz2019nonstationary,Gutierrez2023} (SSR). It was introduced in Refs.~\cite{AGARWAL1989,Agarwal1990} to describe a system of two-level emitters incoherently driven by broadband squeezed vacuum light. 

\begin{figure}
    \centering
    \includegraphics[width=0.48\textwidth]{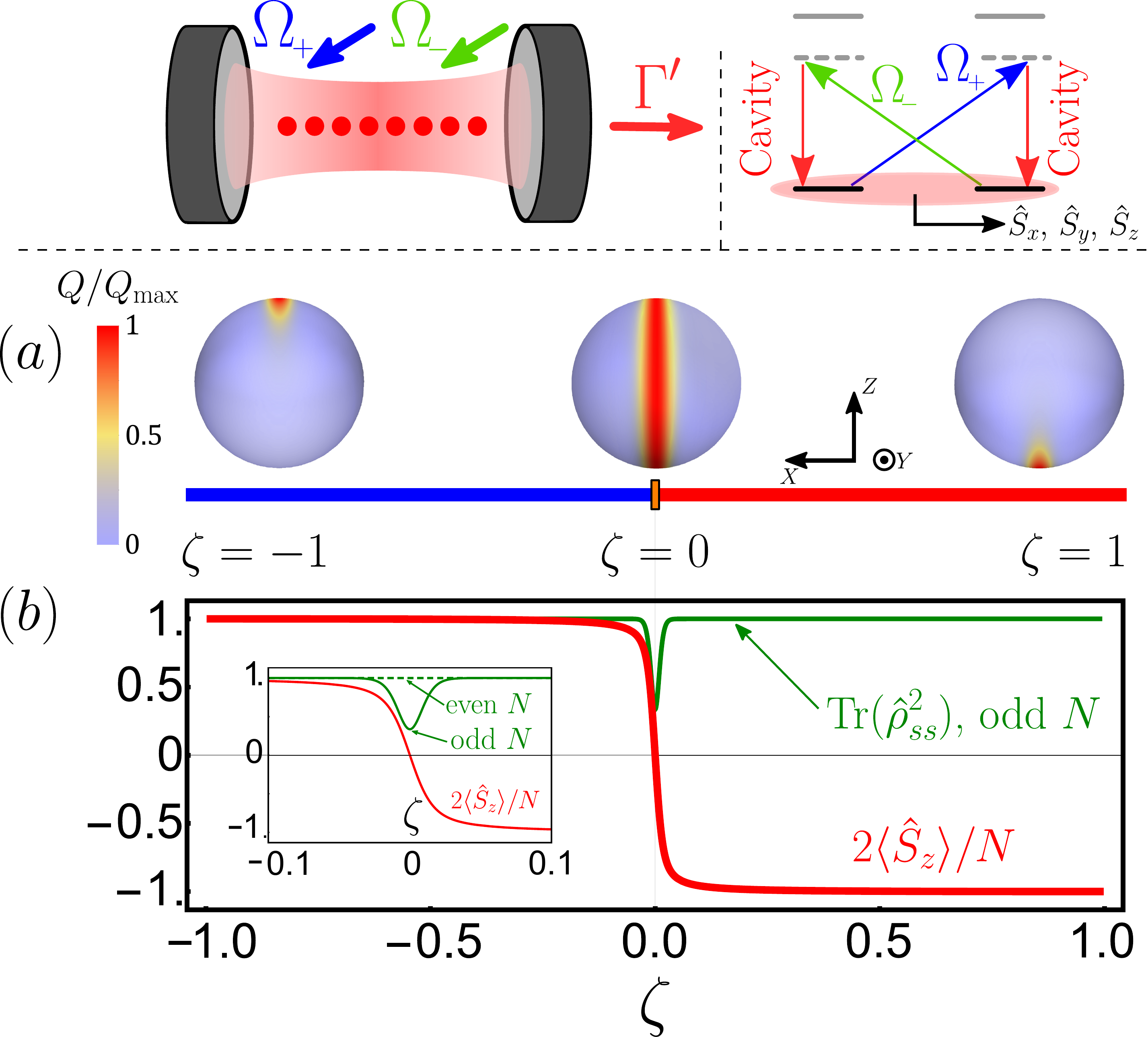}
    \caption{(a) Cavity implementation of squeezed superradiance (SSR) relies on two tones ($\Omega_{\pm}$) driving cavity-assisted two-photon Raman transitions on the two-level system of interest, indicated by the shaded region in the right panel. Light escapes from the cavity with rate $\propto\Gamma'$. (b) Steady state phase diagram of squeezed superradiance as a function of $\zeta$. Spheres show the Husimi distribution of the steady state at $\zeta=-1,0,1$ (from left to right) for $N=101$. (c) Numerical steady state values of $\braket{\hat{S}_z}$ and $\mathrm{Tr}(\hat{\rho}_{\text{ss}}^2)$ (purity) for $N=101$. Inset: close up of the transition region ($\zeta \approx  0$) that showcases the differences between even $N=100$ and odd $N=101$.}
    \label{fig:SDMModel}
\end{figure}

This model can also be engineered inside QED cavities~\cite{DallaTorre2013}, as depicted in Fig.~\ref{fig:SDMModel}(a). In the proposal of Ref.~\cite{DallaTorre2013}, this is achieved by driving two-photon Raman transitions (with Rabi drives $\Omega_{\pm}$) between two degenerate atomic states where one of the Raman legs is provided by a cavity mode. The effects of photon loss through the cavity mirrors (with rate $\propto \Gamma'$) lead to an effective atom-only model described by Eq.~(\ref{eqn:DallaTorre}), where $\zeta\propto (\Omega_+^2-\Omega_-^2)$ is controlled by the imbalance between Rabi drive strengths. In this case, the presence of collective atomic operators $\hat{S}_{x,y,z}$ reflects the fact that the photons escaping the cavity do not carry information about which atoms they were emitted from. This model also provides an example of relaxation towards an entangled dark state by an adequate engineering of dissipation processes~\cite{Kuzmich1997,DallaTorre2013,Groszkowski2022}.

In the thermodynamic limit $N\to\infty$, the model displays a first-order phase transition at $\zeta=0$, as illustrated in Fig.~\ref{fig:SDMModel}(b) and (c):
\begin{itemize}
\item When $-1<\zeta<0$, the steady state is a large spin pointing along the $+z$ direction: $\braket{\hat{S}_z}=N/2$ and $\braket{\hat{S}^-}=0$.
\item When $0<\zeta<1$, the steady state is a large spin pointing along the $-z$ direction: $\braket{\hat{S}_z}=-N/2$ and $\braket{\hat{S}^-}=0$.
\end{itemize}

Far away from the transition point $\zeta=0$, the behaviour of quantum fluctuations about the mean-field state can be described analytically by means of linearization. For completeness, we reproduce here the main results derived from this analysis~\cite{munoz2019nonstationary}.

Without loss of generality we consider $0<\zeta<1$ only~\footnote{The steady state for $\pi/4<\theta<\pi/2$ can be related to the steady state for $0<\theta<\pi/4$ by means of the unitary rotation $\exp(i\pi\hat{S}_x)$}, where the steady state is polarized along the $-z$ direction. This allows us to use the (exact) Holstein-Primakoff~\cite{Holstein1940} representation of spin operators about $-z$
\begin{equation}
    \hat{S}_z=-N/2+\hat{a}^\dagger\hat{a},\hspace{1cm} \hat{S}^-=\Big(\sqrt{N-\hat{a}^\dagger\hat{a}}\Big)\hat{a},
\end{equation}
where $\hat{a}$ is the annihilation operator for an auxiliary boson. The strong polarization along $-z$, valid away from the critical point, is mathematically expressed by the conditions $\braket{\hat{a}^\dagger\hat{a}},\text{Var}(\hat{a}^\dagger\hat{a})^{1/2}\ll N$, which imply that $\hat{S}^-$ can be approximated by $\sqrt{N} \hat{a}$. Within this approximation, Eq.~(\ref{eqn:DallaTorre}) becomes
\begin{equation}
    \dot{\hat{\rho}}=\frac{\Gamma'}{2}\mathcal{D}\big(\hat{x}+i\zeta\hat{p}\big)\hat{\rho},
\end{equation}
where $\hat{x}=(\hat{a}+\hat{a}^\dagger)/\sqrt{2}$ and $\hat{p}=(\hat{a}-\hat{a}^\dagger)/(i\sqrt{2})$ are the bosonic quadrature approximations of $\hat{S}_x$ and $-\hat{S}_y$. The jump operator of the linearized evolution equation is $\hat{x}+i\hat{p}$, which possesses a unique dark state defined by $(\hat{x}+i\zeta\hat{p})\ket{D_\zeta}=0$. At long times, the system evolves towards this state, which satisfies $\braket{\hat{x}}=\braket{\hat{p}}=0$ and
\begin{equation}
    2\braket{\hat{x}^2}=\frac{1}{2\braket{\hat{p}^2}}=\zeta.
\end{equation}
As $\zeta$ approaches $0$, so does $\braket{\hat{x}^2}$, and hence the spin variable $\hat{S}_x$ gets squeezed. However, at the same time $\braket{\hat{p}^2}$ approaches $\infty$, violating the conditions for strong polarization. Solving for the steady state around $\zeta\approx 0$ using this particular instance of the Holstein-Primakoff representation demands that we solve a highly nonlinear bosonic problem and is thus not a convenient route of attack.

The computations within Holstein-Primakoff indicate that, away from the critical point, there is a steady state that is Gaussian and pure for any $N$. However, as shown in Refs.~\cite{AGARWAL1989,Agarwal1990,Groszkowski2022}, a pure steady state exists only when $N$ is even, while the state becomes strongly mixed close to $\zeta=0$ for odd $N$. This is illustrated in the inset of Fig.~\ref{fig:SDMModel}(c) by means of the state purity $\mathrm{Tr}(\hat{\rho}_{\text{ss}}^2)$ for $N=100,101$, but the different behaviours for even/odd $N$ will also be manifested in observables such as variances. For even $N$, Holstein-Primakoff is consistent with the exact result and both describe pure steady states. When $N$ is odd, Holstein-Primakoff and the exact result are in tension, so we conclude that the steady state cannot be pure, but must be only approximately pure. Since the properties of the steady state across the critical point are very sensitive to the parity of $N$~\cite{Groszkowski2022}, we do independent analyses for even and odd $N$. We begin with even $N$, where a proper pure steady state exists.

\subsection{Even N}
\begin{figure*}
    \centering
    \includegraphics[width=0.98\textwidth]{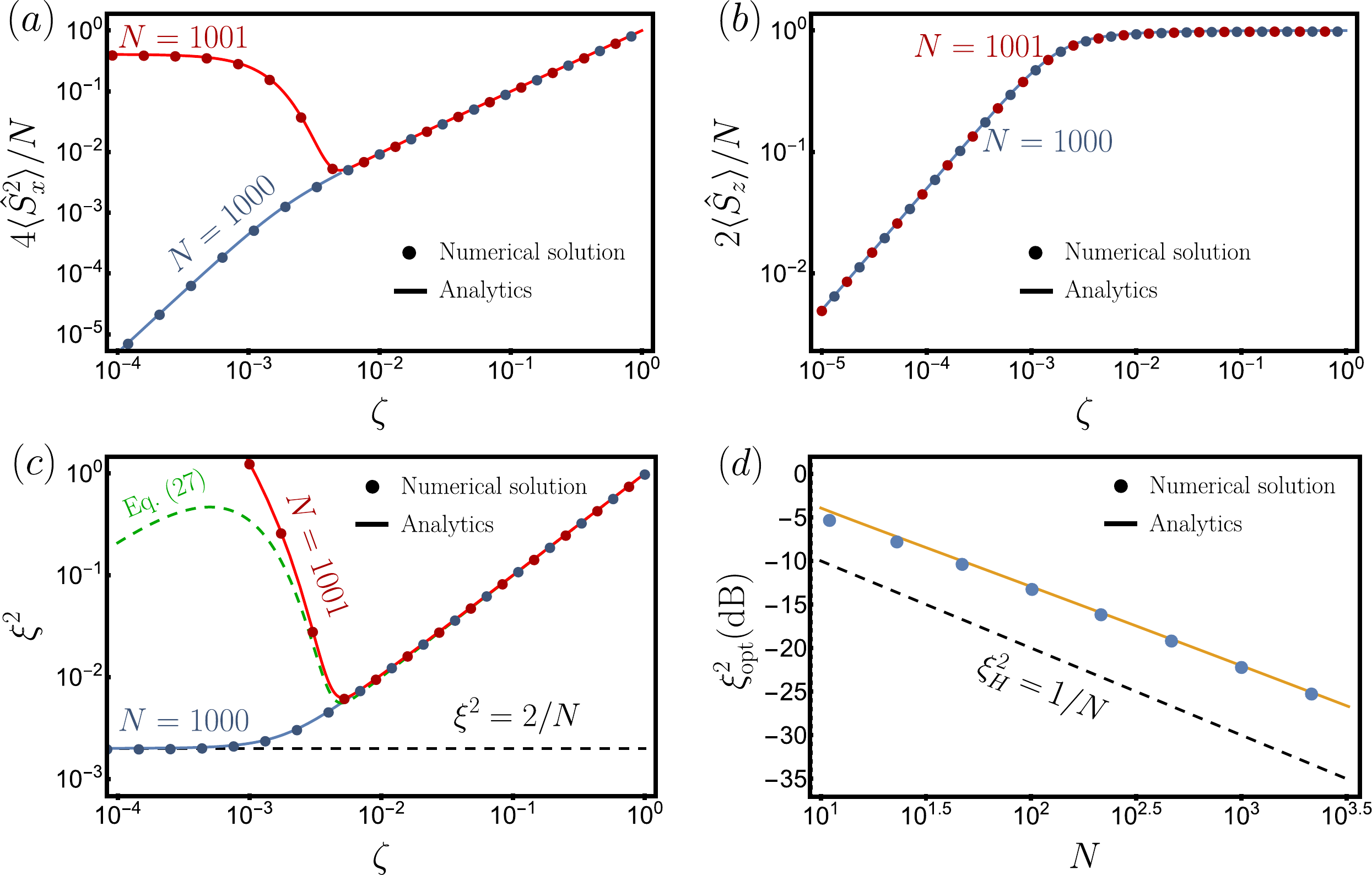}
    \caption{(a) Steady state variance of $\hat{S}_x$ in squeezed superradiance (SSR) as a function of $\zeta$. We show numerical results for $N=1000$ (solid blue) and $N=1001$ (solid red), and the analytical formulas given in Eqs.~(\ref{eqn:EvenNExpectations}) and~(\ref{eqn:OddNSx2}) for $N=1000$ (blue dots) and  $N=1001$ (red dots), respectively. (b) Steady state value of $\hat{S}_z$ is the same for even and odd $N$. The analytical formula is given by Eq.~(\ref{eqn:EvenNExpectations}). (c) Steady state value of $\xi^2$. Analytical formulas are Eq.~(\ref{eqn:SDMEvenNSqueezing}) ($N=1000$, blue dots) and Eq.~(\ref{eqn:OddNSqueezing}) ($N=1001$, red dots). We also show the approximation given by Eq.~(\ref{eqn:SDMMinimumApprox}) (dashed green), which captures correctly the minimum. (d) Optimal squeezing as a function of particle number for $N$ odd. Blue dots are obtained numerically and solid orange line is the first line of Eq.~(\ref{eqn:SDMMinimumSqueezingOddN}).}
    \label{fig:SDMSqueezing}
\end{figure*}

In this case the steady state of the system is pure for all values of $\zeta$~\cite{AGARWAL1989,Agarwal1990}, and is defined by
\begin{equation}\label{eqn:SDMEvenDark}
    (\hat{S}_x-i\zeta\hat{S}_y)\ket{D_\zeta}=0.
\end{equation}
While this equation can be solved exactly, we will instead perform approximations that lead to a more intuitive grasp of the properties of the state $\ket{D_{\zeta}}$ in a way that is also relevant for the odd $N$ case. We do this by taking advantage of the fact that the steady state is highly concentrated around $\hat{S}_x=0$ [see Fig.~\ref{fig:SDMModel}(b)] and using another Holstein-Primakoff representation, but this time about the $+x$ direction:
\begin{equation}
    \hat{S}_x=\frac{N}{2}-\hat{b}^\dagger\hat{b},\hspace{1cm}\hat{S}_z-i\hat{S}_y=-\Big(\sqrt{N-\hat{b}^\dagger\hat{b}}\Big)\hat{b},
\end{equation}
where $\hat{b}$ is a different auxiliary boson. Since $\braket{\hat{S}_x}\approx 0$ in the steady state, the boson $\hat{b}$ is highly excited, with $\braket{\hat{b}^\dagger\hat{b}}\approx N/2$. Nevertheless, the fluctuations in $\hat{S}_x$ are of size $\sim\sqrt{N}$ and translate into $\sim\sqrt{N}$ fluctuations in boson excitation, which are thus very small relative to its extensive occupation. Under these conditions, it becomes advantageous to use the number-phase representation of bosons~\cite{Susskind1964} as the starting point of our approximations: 
\begin{equation}
     \hat{b}=e^{i\hat{\phi}}\,\sqrt{\hat{n}},
\end{equation}
where $\hat{n}=\hat{b}^\dagger\hat{b}$ is the number operator, $e^{i\hat{\phi}}$ reduces boson occupation with unit amplitude~\footnote{There are subtleties around the boson vaccuum, but they are of no relevance here due to the macroscopic $\sim N$ boson excitation}, and they satisfy the commutation relation $[\hat{n},e^{i\hat{\phi}}]=-e^{i\hat{\phi}}$. To lowest order in $1/N$ we can replace $\hat{n}\approx N/2$ in the equations for $\hat{S}_{y,z}$, so that we have
\begin{equation}\label{eqn:SDMXHolstein}
    \hat{S}_x=-\delta\hat{n},\hspace{0.35cm} \hat{S}_z\approx-\frac{N}{2}\cos\hat{\phi},\hspace{0.35cm} \hat{S}_y\approx\frac{N}{2}\sin\hat{\phi},
\end{equation}
where $\delta\hat{n}\equiv \hat{n}-N/2$ measures the excitation number with respect to its macroscopic occupation. The pure steady state condition is then reframed as
\begin{equation}
    \bigg(\delta\hat{n}+\frac{i\zeta N}{2}\sin\hat{\phi}\bigg)\ket{D_{\zeta}}=0.
\end{equation}
This equation can be solved by resorting to wavefunctions in the $\hat{\phi}$ representation: $D_{\zeta}(\phi)=\braket{\phi|D_{\zeta}}$, in which $\delta\hat{n}$ becomes $i\partial_\phi$. Solving the ensuing first order differential equation leads to the following (unnormalized) wavefunction for the dark state
\begin{equation}
    D_{\zeta}(\phi)=e^{\frac{\zeta N}{2}\cos\phi}.
\end{equation}
Away from the critical point $\zeta N$ is much larger than $1$, so the wavefunction is concentrated around $\phi=0$, indicating that the state is polarized along $-z$. Instead, when $\zeta N\lesssim 1$, the wavefunction is distributed along a full circle, indicating proximity to a $\hat{S}_x$ eigenstate. Using the explicit form for $D_{\zeta}(\phi)$ we can calculate observables such as 
\begin{equation}\label{eqn:EvenNExpectations}
    \braket{\hat{S}_z}=-\frac{N}{2}\frac{I_1(\zeta N)}{I_0(\zeta N)},\hspace{1cm} \braket{\hat{S}_x^2}=\frac{\zeta N}{4}\frac{I_1(\zeta N)}{I_0(\zeta N)},
\end{equation}
where $I_k$ is the $k^{\text{th}}$ modified Bessel function of the first kind. As $\zeta\to 0$, both $\braket{\hat{S}_z}$ and $\braket{\hat{S}_x^2}$ approach 0 since the dark state approaches the $\hat{S}_x=0$ eigenstate close to the critical point. However, the spin squeezing parameter~\cite{Wineland1992}
\begin{equation}\label{eqn:SDMEvenNSqueezing}
    \xi^2\equiv N\frac{\text{Var}(\hat{S}_x)}{\braket{\hat{S}_z}^2}=\frac{1}{N}\bigg[\zeta N\frac{I_0(\zeta N)}{I_1(\zeta N)}\bigg],
\end{equation}
approaches the finite value $2/N$. We benchmark these results by numerically computing $\braket{\hat{S}_z}$, $\braket{\hat{S}_x^2}$ and $\xi^2$ as a function of $\zeta$ for $N=1000$ using Eq.~(\ref{eqn:SDMEvenDark}), and show the results in Fig.~\ref{fig:SDMSqueezing}. We find excellent agreement between the numerical calculations and our analytical expressions.
\subsection{Odd N}\label{sec:subOddN}
The steady state is not truly pure in this situation, but can still be written exactly~\cite{AGARWAL1989,Agarwal1990,Groszkowski2022}
\begin{equation}\label{eqn:SDMOddDark}
    \hat{\rho}_{\text{ss}}^{\text{I}}\equiv\bigg(\frac{1}{\hat{S}_x-i\zeta\hat{S}_y}\bigg)\bigg(\frac{1}{\hat{S}_x+i\zeta\hat{S}_y}\bigg)
\end{equation}
up to a normalization factor, which we will calculate later. For this discussion it is convenient to introduce the spectrum of the steady state: $\hat{\rho}_{\text{ss}}^{\text{I}}\ket{\lambda_k}=\lambda_k\ket{\lambda_k}$, where $k$ indexes eigenvalues in order of decreasing magnitude. Equivalently,
\begin{equation}
    (\hat{S}_x+i\zeta\hat{S}_y)(\hat{S}_x-i\zeta\hat{S}_y)\ket{\lambda_k}=\frac{1}{\lambda_k}\ket{\lambda_k}.
\end{equation}
If $N$ were even, then $1/\lambda_0$ would be 0, $\lambda_0$ would be $\infty$ (while all other $\lambda_k$ would be finite) and the weight of $\ket{\lambda_0}$ in the steady state would be infinitely larger than the weight of the other eigenstates of $\mathcal{I}^\dagger\mathcal{I}$. The steady state would then be exactly pure. When $N$ is odd this is no longer the case and the rest of eigenstates may contribute to physical observables. Nevertheless, the Holstein-Primakoff linearization about $-z$ indicates that the steady state should be approximately pure when $\zeta$ is not close to 0, i.e. $\lambda_0$ should be very large, though not $\infty$.

Since the state is still squeezed along $\hat{S}_x$, as in the even $N$ case, we use the Holstein-Primakoff replacements about $+x$ given in Eq.~(\ref{eqn:SDMXHolstein}) to arrive at
\begin{align}\begin{split}\label{eqn:OddNSSSpectrum}
    \bigg(\delta\hat{n}-\frac{i\zeta N\sin\hat{\phi}}{2}\bigg)\bigg(\delta\hat{n}+\frac{i\zeta N\sin\hat{\phi}}{2}\bigg)\ket{\lambda_k}&=\frac{\ket{\lambda_k}}{\lambda_k}.
\end{split}\end{align}
In fact, $\lambda_0$ can be calculated analytically (see Appendix~\ref{sec:AppFirst})
\begin{equation}\label{eqn:OddNEigenvalue}
    \lambda_0=\pi^2\big[I_0(\zeta N)\big]^2,
\end{equation}
which is exponentially large in $\zeta N$. The wavefunction of $\ket{\lambda_0}$ is approximately the same wavefunction as that of $\ket{D_\zeta}$, namely $\braket{\phi|\lambda_0}\propto \exp(\zeta N\cos\phi/2)$, so that expectation values with respect to this state can be imported directly from Eq.~(\ref{eqn:EvenNExpectations}). However, to calculate the steady state observables we need to perform a sum over the contributions of all eigenstates of $\hat{\rho}_{\text{ss}}^{\text{I}}$, not just the first one. For example, the normalization factor is
\begin{equation}
    \mathrm{Tr}(\hat{\rho}_{\text{ss}}^{\text{I}})=\lambda_0+\sum_{k=1}^N\lambda_k.
\end{equation}
We know $\lambda_0$ can be large [of size $e^{2\zeta N}$] so the relevant question is whether the other eigenvalues introduce some extra $N$ dependence. We can get a rough idea of their contribution by considering $\lambda_k$ in Eq.~(\ref{eqn:OddNSSSpectrum}) for $k$ large, meaning $\lambda_k$ small. The $\sin(\phi)$ terms are of size $1$ when $\zeta N\sim 1$ but the eigenvalues are large ($\lambda_k^{-1}\gg 1$), so they are mostly determined by $\delta\hat{n}=-\hat{S}_x$ and the eigenstates are, to a good approximation, $\hat{S}_x$ eigenstates. The spectrum of $\hat{S}_x$ is comprised of half-integers and can be identified with $k+1/2$. Thus, for large $k$, Eq.~(\ref{eqn:OddNSSSpectrum}) becomes $\hat{S}_x\ket{\lambda_k}\approx \lambda_k^{-1}\ket{\lambda_k}$ and we see that $\lambda_k\sim k^{-2}$. Because of this, the sum quickly converges to an $N$ independent value and we can approximate $\mathrm{Tr}(\hat{\rho}_{\text{ss}}^{\text{I}})\approx \lambda_0$. Similarly
\begin{equation}
    \mathrm{Tr}(\hat{\rho}_{\text{ss}}^{\text{I}}\hat{S}_z)=\lambda_0\braket{\lambda_0|\hat{S}_z|\lambda_0}+\sum_{k=1}^N\lambda_k\braket{\lambda_k|\hat{S}_z|\lambda_k}.
\end{equation}
Since the $\ket{\lambda_k}$ approach $\hat{S}_x$ eigenstates, $\braket{\lambda_k|\hat{S}_z|\lambda_k}$ approaches $0$ as $k$ increases and the sum converges even faster than the sum for the normalization. Thus $\mathrm{Tr}(\hat{\rho}_{\text{ss}}^{\text{I}}\hat{S}_z)\approx \lambda_0\braket{\lambda_0|\hat{S}_z|\lambda_0}$ and
\begin{equation}
    \braket{\hat{S}_z}=\frac{\mathrm{Tr}(\hat{\rho}_{\text{ss}}^{\text{I}}\hat{S}_z)}{\mathrm{Tr}(\hat{\rho}_{\text{ss}}^{\text{I}})}\approx -\frac{N}{2}\frac{I_1(\zeta N)}{I_0(\zeta N)},
\end{equation}
using Eq.~(\ref{eqn:EvenNExpectations}). The profile for $\braket{\hat{S}_z}$ as a function of $\zeta$ is therefore the same as the one for the even $N$ case, which is illustrated in Fig.~\ref{fig:SDMSqueezing}(b). Finally,
\begin{equation}\label{eqn:SDMBulkSum}
    \mathrm{Tr}(\hat{\rho}_{\text{ss}}^{\text{I}}\hat{S}_x^2)=\lambda_0\braket{\lambda_0|\hat{S}_x^2|\lambda_0}+\sum_{k=1}^N\lambda_k\braket{\lambda_k|\hat{S}_x^2|\lambda_k}.
\end{equation}
In this case, $\braket{\lambda_k|\hat{S}_x^2|\lambda_k}$ increases as $k^2$ as $k$ increases, canceling the $k^{-2}$ from $\lambda_k$. Thus, the sum contributes a term that scales like $N$. The exact prefactor can be calculated using semiclassical analysis (see Appendix~\ref{sec:AppBulk}) and leads to
\begin{equation}\label{eqn:OddNSx2}
    \braket{\hat{S}_x^2}\approx \frac{\zeta N}{4}\frac{I_1(\zeta N)}{I_0(\zeta N)}+\bigg(\frac{N}{1+\zeta}\bigg)\lambda_0^{-1},
\end{equation}
where $\braket{\hat{S}_x^2}=\mathrm{Tr}(\hat{\rho}_{\text{ss}}^{\text{I}}\hat{S}_x^2)/\mathrm{Tr}(\hat{\rho}_{\text{ss}}^{\text{I}})\approx \lambda_0^{-1}\mathrm{Tr}(\hat{\rho}_{\text{ss}}^{\text{I}}\hat{S}_x^2)$ and we have used Eq.~(\ref{eqn:EvenNExpectations}) to replace $\braket{\lambda_0|\hat{S}_x^2|\lambda_0}$. In contrast to the even $N$ case, when $\zeta\ll 1/N$ the $\hat{S}_x$ variance approaches a constant nonzero value (though still below the quantum projection noise value $N/4$), determined by all the eigenstates $\ket{\lambda_k}$, even if the contribution from $\ket{\lambda_0}$ vanishes. Since the contrast $\braket{\hat{S}_z}$ does go to 0, this means that the squeezing parameter should have a minimum at a finite value of $\zeta$. Putting these results together leads to the following functional form for the squeezing parameter
\begin{equation}\label{eqn:OddNSqueezing}
    \xi^2\approx\zeta\bigg[\frac{I_0(\zeta N)}{I_1(\zeta N)}\bigg]+\frac{4}{\pi^2 I_1(\zeta N)^2},
\end{equation}
where we have set $\zeta\approx 0$ in all expressions except in the arguments of the Bessel functions (where $\zeta$ is multiplied by $N$). The first term comes from the $\ket{\lambda_0}$ state, while the second one comes from the other eigenstates $\ket{\lambda_k}$. We benchmark these results by numerically calculating $\braket{\hat{S}_z}$, $\braket{\hat{S}_x^2}$ and $\xi^2$ from the steady state density matrix $\hat{\rho}_{\text{ss}}^{\text{I}}=1/(\mathcal{I}^\dagger\mathcal{I})$ for $N=1001$ and comparing them against Eq.~(\ref{eqn:OddNSx2}) and Eq.~(\ref{eqn:OddNSqueezing}) in Fig.~\ref{fig:SDMSqueezing}. We find good agreement, and showcase the presence of a minimum $\xi^2$. 

To obtain an analytical estimate of the minimum, we further assume that $\zeta_{\text{min}}N\gtrsim 1$ so that we can approximate $\xi^2$ as
\begin{equation}\label{eqn:SDMMinimumApprox}
    \xi^2=\zeta+\frac{8\zeta N}{\pi}e^{-2\zeta N}.
\end{equation}
This approximation captures correctly the minimum, as shown in Fig.~\ref{fig:SDMSqueezing}(c), but misses the behaviour at smaller $\zeta N$. We minimize this expression with respect to $\zeta$ to obtain
\begin{align}\begin{split}\label{eqn:SDMMinimumSqueezingOddNzeta}
    \zeta_{\text{min}}N&=\frac{1}{2}\bigg[1-W_{-1}\bigg(-\frac{\pi e}{8N}\bigg)\bigg]\\
    \zeta_{\text{min}}N&\approx \frac{1}{2}\bigg[\ln\bigg(\frac{8N}{\pi}\bigg)+\ln\ln\bigg(\frac{8N}{\pi e}\bigg)\bigg]\\
    &\hspace{3.5cm}+O\bigg(\frac{\ln\ln N}{\ln N}\bigg),
\end{split}\end{align}
where $W_{-1}$ is the $-1$ branch of the Lambert W function, and the second line is the result of an expansion of $W_{-1}$  about $N=\infty$. Notice that as $N$ increases so does $\zeta_{\text{min}}N$, which guarantees the validity of the approximation $\zeta_{\text{min}}N\gtrsim 1$ for large $N$. Similarly
\begin{align}\begin{split}\label{eqn:SDMMinimumSqueezingOddN}
    \xi^2_{\text{min}}&=\zeta_{\text{min}}\bigg(1+\frac{1}{2\zeta_{\text{min}}N-1}\bigg)\\
     \xi^2_{\text{min}}&\approx \frac{1}{2N}\bigg[\ln\bigg(\frac{8N e}{\pi }\bigg)+\ln\ln\bigg(\frac{8N}{\pi e}\bigg)\bigg].
\end{split}\end{align}
Thus, in the case of odd $N$ there is a logarithmic correction to the scaling of the minimum squeezing $\xi^2$. While these corrections are mild, they can create confusion about finite size scalings if a power law dependence on $N$ is fitted naively. For example, fitting a power law for $\xi^2_{\text{min}}$ obtained numerically between $N=10^2$ and $N=10^4$ gives a behaviour consistent with $N^{-0.9}$. Furthermore, cleanly observing the logarithmic behaviour numerically is very challenging since corrections are of relative size $\log\log N/\log N$, which is only $0.07$ even for $N=10^{23}$. Instead, the first lines of Eqs.~(\ref{eqn:SDMMinimumSqueezingOddNzeta}) and~(\ref{eqn:SDMMinimumSqueezingOddN}) are more accurate expressions [as shown in Fig.~\ref{fig:SDMSqueezing}(d) for the optimal $\xi^2$], with relative corrections of size $N^{-1}$. Note also that this logarithmic behaviour is a consequence of the minimization process. If we had considered fixed $\zeta N$, Eq.~(\ref{eqn:OddNSqueezing}) indicates that $\xi^2$ reaches a constant value as $N$ is increased.

\section{Model II: Driven Superradiance}\label{sec:CRF}
\begin{figure}
    \centering
    \includegraphics[width=0.48\textwidth]{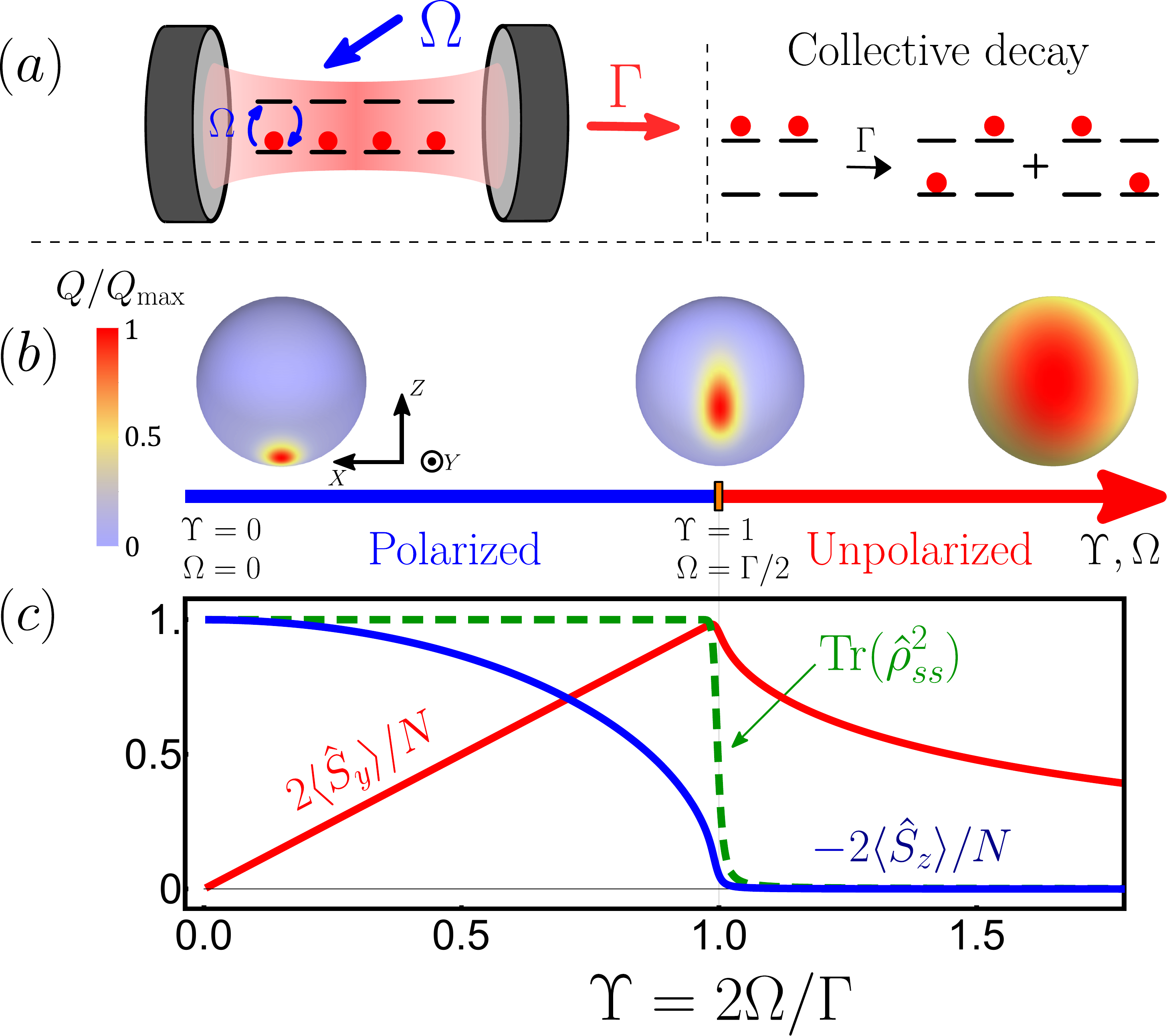}
    \caption{(a) Implementation of driven superradiance in an optical cavity with a Rabi drive ($\Omega$) and collective emission ($\Gamma$) through the cavity mirrors, which creates correlations between atoms. (b) Steady state phase diagram of driven superradiance as a function of $\Upsilon=2\Omega/\Gamma$. Spheres show the Husimi distribution of the steady state at $\Upsilon=0.75, 1, 2$ (from left to right) for $N=100$. (c) Numerical steady state values of $\braket{\hat{S}_{y}}$, $\braket{\hat{S}_z}$ and $\mathrm{Tr}(\hat{\rho}_{\text{ss}}^2)$ (purity) for $N=1000$. }
    \label{fig:CRFModel}
\end{figure}

We now switch to the second model, given by Eq.~(\ref{eqn:CollectiveResonanceFluoresence}), which we reproduce here for reference
\begin{equation}\label{eqn:CollectiveResonanceFluoresenceII}
    \partial_t\hat{\rho}=-i\big[\Omega\hat{S}_x,\hat{\rho}\big]+\frac{\Gamma}{N} \mathcal{D}(\hat{S}^-)\hat{\rho},
\end{equation}
and whose (unnormalized) steady state can also be written exactly~\cite{@Puri1979,Kilin1980}
\begin{equation}\label{eqn:CRFSteadyState}
    \hat{\rho}_{\text{ss}}^{\text{II}}=\bigg(\frac{1}{\hat{S}^-+\frac{i N\Upsilon}{2}}\bigg)\bigg(\frac{1}{\hat{S}^+-\frac{i N\Upsilon}{2}}\bigg),
\end{equation}
where $\Upsilon=2\Omega/\Gamma$. We will refer to this model as driven superradiance, for clarity. The model has two ingredients. One is the external drive $\Omega\hat{S}_x$, which induces Rabi flopping between the two atomic levels we are considering. In the context of an optical cavity, it can be engineered by shining the atoms with laser light resonant with the two-level transition [see Fig.~\ref{fig:CRFModel}(a), left]. The second ingredient is collective decay, mathematically described by the jump operator $\sqrt{\Gamma/N}\hat{S}^-$. This effective decay process is a consequence of photon leakage through the cavity mirrors, whereby atomic excitations are transformed into intracavity photons that then quickly escape the system. As in squeezed superradiance, these photons do not carry information about which atom they were emitted from, so the decay process is collective [see Fig.~\ref{fig:CRFModel}(a), right]. The effective atomic description in terms of $\hat{S}^-$, which neglects intracavity photon dynamics, arises after adiabatic elimination of the cavity degree of freedom, which is a valid procedure when the lifetime of a photon inside the cavity is much shorter than any other relevant timescale~\cite{Bonifacio1971}. Equation~(\ref{eqn:CollectiveResonanceFluoresenceII}) also describes a simplified instance of the phenomenon known as cooperative resonance fluorescence in the limit where single particle emission into free space is neglected~\cite{Carmichael_1980,Drummondo1980,Walls1980}.

In the thermodynamic limit, driven superradiance displays a continuous steady state phase transition as a function of $\Upsilon=2\Omega/\Gamma$, as shown in Fig.~\ref{fig:CRFModel}~\cite{Carmichael_1980,Drummondo1980,Walls1980}: 
\begin{itemize}
    \item When $\Upsilon<1$, the steady state is a highly pure polarized state pointing along 
    \begin{align}\begin{split}\label{eqn:CRFMeanFieldObservables}
    \Big[\langle\hat{S}_x\rangle,\langle\hat{S}_y\rangle,\langle\hat{S}_z\rangle\Big]=\frac{N}{2}\Big[0,\Upsilon,-\sqrt{1-\Upsilon^2}\Big]
    \end{split}\end{align}
    on the southern hemisphere of the Bloch sphere. It arises from the equilibration between superradiant decay $\Gamma$, which pulls the state towards the south pole, and the drive $\Omega$, which rotates the state away from the south pole.
    \item When $\Upsilon>1$ the drive cannot be equilibrated by superradiant decay, and the steady state arises instead from a slow collective-decay-induced diffusion process of the classical mean-field trajectories. As a consequence, the state is strongly mixed [as measured by the purity $\mathrm{Tr}(\hat{\rho}_{\text{ss}}^2)$, see Fig.~\ref{fig:CRFModel}(c)], and its Husimi distribution on the Bloch sphere is very diffuse [see Fig.~\ref{fig:CRFModel}(b)]. 
    In this regime, the state of the system can be described by means of a classical distribution on the sphere, 
    \begin{equation}\label{eqn:CRFClassicalDistribution}
    \rho_{\text{ss}}^{\text{cl}}(\theta,\phi)=\frac{(N/2)^{-2}}{|\sin\theta\, e^{-i\phi}+i\Upsilon|^2},
    \end{equation}
    which is parameterized in terms of the spherical coordinate angles $\theta$ and $\phi$, and is obtained by replacing the quantum operator $\hat{S}^-$ with $N\sin\theta e^{-i\phi}/2$ in Eq.~(\ref{eqn:CRFSteadyState}). The leading contribution (in a $1/N$ expansion) to observables is obtained by using this replacement, together with $\hat{S}_z\to N\cos\theta/2$, and integrating with respect to the surface measure $N\sin\theta\,d\theta\,d\phi/4\pi$. Using this prescription, we get 
    \begin{equation}
    \braket{\hat{S}_y}=\frac{N}{2\Upsilon}\Bigg[\Upsilon^2-\frac{\sqrt{\Upsilon^2-1}}{\arcsin\big(\Upsilon^{-1}\big)}\Bigg],
    \end{equation}
    in agreement with Refs.~\cite{Carmichael_1980,Drummondo1980}, which obtain the same result through other methods.
\end{itemize}
In contrast to SSR, this transition is continuous and can be related to notions of symmetry-breaking, thus establishing a connection to second-order phase transitions in equilibrium~\cite{Hannukainen2018,Link2019}. Experimental access to this behaviour can be achieved by changing the intensity of the laser that creates the Rabi drive $\Omega$ [Fig.~\ref{fig:CRFModel}(a)], which controls the size of $\Upsilon$.

The properties of the system near the transition point are better understood by analyzing fluctuations about the mean-field state in the polarized phase ($\Upsilon<1$), so we reproduce these results here~\cite{Carmichael_1980,Drummondo1980,Walls1980}. The analysis is made easier by rewriting Eq.~(\ref{eqn:CollectiveResonanceFluoresenceII}) as
\begin{equation}\label{eqn:CRFMasterEquation}
    \dot{\hat{\rho}}=\frac{\Gamma}{N}\mathcal{D}\bigg(\hat{S}^-+\frac{iN\Upsilon}{2}\bigg)\hat{\rho}.
\end{equation}

When $\Upsilon<1$, the steady state is polarized along a direction on the Bloch sphere that lies on the $yz$ plane. If we denote the angle between this direction and the $-z$ axis as $\alpha$, then the Bloch vector will satisfy
\begin{equation}
    \braket{\hat{S}_x}=0,\hspace{0.3cm}\braket{\hat{S}_y}=\frac{N}{2}\sin\alpha,\hspace{0.3cm} \braket{\hat{S}_z}=-\frac{N}{2}\cos\alpha.
\end{equation}
Mean field analysis [Eq.~(\ref{eqn:CRFMeanFieldObservables})] indicates that $\sin\alpha=\Upsilon$ and hence this polarized state exists only for $\Upsilon\leq1$. 

The steady state will not be a coherent state, and quantum fluctuations about the mean field direction will suffer small modifications. This can be cleanly analyzed in the large $N$ limit by doing a Holstein-Primakoff linearization about the polarization direction. This is implemented by defining rotated spin operators $\hat{S}_{y,z}'$ implicitly through
\begin{align}
    \begin{split}\label{eqn:CRFRotation}
        \hat{S}_y&=\hat{S}_y'\cos\alpha-\hat{S}_z'\sin\alpha\\
        \hat{S}_z&=\hat{S}_z'\cos\alpha+\hat{S}_y'\sin\alpha.
    \end{split}
\end{align}
In this rotated frame, $\braket{\hat{S}_z'}=-N/2$ and $\braket{\hat{S}_y'}=0$. The Holstein-Primakoff approximation then becomes
\begin{equation}\label{eqn:CRFLinearization}
    \hat{S}_z'\approx -\frac{N}{2},\hspace{0.4cm}\hat{S}_y'\approx -\hat{p}\sqrt{\frac{N}{2}},\hspace{0.4cm} \hat{S}_x\approx \hat{x}\sqrt{\frac{N}{2}},
\end{equation}
where $(\hat{x},\hat{p})$ are the canonically conjugate quadratures of the Holstein-Primakoff auxiliary boson. Plugging this into the master equation and enforcing the mean field equilibration condition ($\Upsilon=\sin\alpha$) leads to
\begin{equation}
    \hat{\rho}=\frac{\Gamma}{2}\mathcal{D}\big(\hat{x}+i\hat{p}\cos\alpha\big)\hat{\rho}.
\end{equation}
The jump operator $\hat{x}+i\hat{p}\cos\alpha$ possesses a dark state that satisfies $\braket{\hat{x}}=\braket{\hat{p}}=0$ and
\begin{equation}\label{eqn:CRFsqueezingbelow}
    2\braket{\hat{x}^2}=\frac{1}{2\braket{\hat{p}^2}}=\cos\alpha=\sqrt{1-\Upsilon^2}.
\end{equation}
This indicates that the $\hat{S}_x\propto \hat{x}$ variable is squeezed, with squeezing parameter $\xi^2=\cos\alpha$. As $\Upsilon\to 1$, $\braket{\hat{p}^2}$ grows without bound, violating the conditions for linearization [Eq.~(\ref{eqn:CRFLinearization})]. Away from criticality, this Gaussian pure state is a very good description of the steady state in the large $N$ limit.

Close to the transition point, the system switches from a highly pure to a highly mixed state in a very narrow parameter region. We can expect some of the phenomenology of section \ref{sec:subOddN} to be applicable, with the relative weight of the first eigenstate of the steady state density matrix dominating when $\Upsilon<1$, but becoming comparable to the rest of eigenstates as $\Upsilon\to 1$. We analyze this in detail in the next subsection.
\subsection{Critical steady state}
We begin from the closed form solution for the (unnormalized) steady state
\begin{equation}\label{eqn:CRFSteadyStateb}
    \hat{\rho}_{\text{ss}}^{\text{II}}=\bigg(\frac{1}{\hat{S}^-+\frac{i N\Upsilon}{2}}\bigg)\bigg(\frac{1}{\hat{S}^+-\frac{i N\Upsilon}{2}}\bigg),
\end{equation}
and look at eigenstates of $\hat{\rho}_{\text{ss}}^{\text{II}}\ket{\mu_k}=\mu_k\ket{\mu_k}$ to investigate their relative weight across the transition and their contribution to various physical observables. These states satisfy
\begin{equation}\label{eqn:CRFEigs}
    \bigg(\hat{S}^+-\frac{i N\Upsilon}{2}\bigg)\bigg(\hat{S}^-+\frac{i N\Upsilon}{2}\bigg)\ket{\mu_k}=\frac{1}{\mu_k}\ket{\mu_k}.
\end{equation}
We can expect that $\mu_0$ is very large when $\Upsilon<1$ to account for the presence of a pure steady state within the Holstein-Primakoff linearization, and therefore we need to calculate its contribution independently from the rest of eigenstates. Close to the transition ($\Upsilon\approx 1$ and $\alpha\approx \pi/2$) the state is still polarized along $+y$, but fluctuations acquire a very nonlinear character and noise distributions are no longer ellipses.
\subsubsection{First eigenstate}
We can hope that keeping higher order terms in the Holstein-Primakoff expansion may be enough to characterize the properties of $\ket{\mu_0}$ and the first few $\ket{\mu_k}$. To proceed along these lines, we fix $\alpha=\pi/2$ in Eq.~(\ref{eqn:CRFRotation}) and let 
\begin{equation}
    \hat{S}_z'=\frac{N}{2}-\hat{a}^\dagger\hat{a}\approx \frac{N}{2}-\frac{\hat{p}^2}{2}.
\end{equation}
We neglect the contributions of $\hat{x}^2$ and $1/2$ to $\hat{a}^\dagger\hat{a}$ because $\hat{p}^2$ is much larger [$\hat{S}_y\sim\hat{p}$ is strongly antisqueezed, see Eq.~(\ref{eqn:CRFsqueezingbelow})]. Then Eq.~(\ref{eqn:CRFEigs}) becomes
\begin{align}\begin{split}
   \bigg[\hat{x}-\frac{i(\hat{p}^2+N\delta\Upsilon)}{\sqrt{2N}}\bigg]\bigg[\hat{x}+\frac{i(\hat{p}^2+N\delta\Upsilon)}{\sqrt{2N}}\bigg]\ket{\mu_k}=\frac{2/N}{\mu_k}\ket{\mu_k}, 
\end{split}\end{align}
where $\delta\Upsilon=\Upsilon-1$ indicates the deviation from the critical point. This is a very small quantity by assumption, and its $N$ scaling must be determined self-consistently. The competition between $\hat{x}$ and the extra $\hat{p}^2$ terms, which come from the higher order terms in Holstein-Primakoff, should stabilize the state and will give it a finite $N$ dependent variance (see~\cite{Liberti2006,Liberti2010} and Supplementary Material of Ref.~\cite{Titum2020}). This is made manifest by introducing canonically re-scaled variables 
\begin{equation}\label{eqn:Rescaling}
    \hat{y}=(2N)^{\frac{1}{6}}\hat{x},\hspace{1cm}\hat{q}=(2N)^{-\frac{1}{6}}\hat{p}.
\end{equation}
To make sure that the terms with $\delta\Upsilon$ remain of the same size as the rest we thus need to scale it as $\delta\Upsilon\equiv (2N)^{-\frac{2}{3}}(2\eta)$, leading to
\begin{align}
    \begin{split}\label{eqn:CRFBosonEquation}
        \big(\hat{y}-i\hat{q}^2-i\eta\big)\big(\hat{y}+i\hat{q}^2+i\eta\big)\ket{\mu_k}=\frac{1}{\tilde{\mu}_k}\ket{\mu_k},
    \end{split}
\end{align}
where $\tilde{\mu}_k=(N/4)^{2/3}\mu_k$ and all the $N$ dependence has now been pushed to various prefactors.  Self-similarity requires that we fix $\eta$ and scale $\delta\Upsilon \sim N^{-2/3}$ when we increase $N$.

Equation (\ref{eqn:CRFBosonEquation}) describes an anharmonic oscillator. Its lowest eigenstate is non-degenerate and its corresponding eigenvalue is close to 0 and can be determined analytically when $\eta\lesssim -1$ (see Appendix \ref{sec:AppFirstCRF}), 
\begin{align}\begin{split}\label{eqn:CRFLowestEigenvalue}
    \tilde{\mu}_0&\approx \bigg[\int_0^\infty e^{-2\eta q-2q^3/3}\,dq\bigg]^2\approx
    \frac{\pi \exp\big(8|\eta|^{3/2}/3\big)}{2\sqrt{|\eta|}}.
\end{split}\end{align}
As expected, this eigenvalue is very large when $\eta$ is large (since $\eta \sim N^{2/3}\delta\Upsilon $). Since the rest of eigenvalues of Eq.~(\ref{eqn:CRFBosonEquation}) do not have the same kind of exponential dependence with $\eta$, this guarantees that the steady state is essentially pure in this regime. Within the same approximation ($\eta\lesssim -1$), the variance of $\hat{y}$ is $\braket{\mu_0|\hat{y}^2|\mu_0}\approx \sqrt{|\eta|}$, a result that we will eventually use to estimate the squeezing. 
\subsubsection{Steady state observables}
To compute observables we also need some information about the excited states of Eq.~(\ref{eqn:CRFBosonEquation}). At high excitation we can neglect $\eta$, which reduces solving Eq.~(\ref{eqn:CRFBosonEquation}) to finding the energy eigenstates of a quartic oscillator. For highly excited states these can be estimated using semiclassical analysis. This procedure indicates that $\tilde{\mu}_k\sim k^{-4/3}$, $\braket{\hat{y}^2}_{\mu_k}\sim k^{4/3}$ and $\braket{\hat{q}^2}_{\mu_k}\sim k^{2/3}$, where $\braket{\,}_{\mu_k}$ is the expectation value with respect to the state $\ket{\mu_k}$. Using these, we can calculate expectation values of the full steady state. We begin with the normalization factor
\begin{equation}
    \mathrm{Tr}(\hat{\rho}_{\text{ss}}^{\text{II}})=\bigg(\frac{N}{4}\bigg)^{-2/3}\bigg(\tilde{\mu}_0+\sum_{k=1}^N\tilde{\mu}_k\bigg)
\end{equation}
The sum $\sum_k k^{-4/3}$ converges in the limit $N\to \infty$ so we keep only the contribution of $\mu_0$. For the $\hat{S}_x$ variance, which is related to $\hat{y}$,
\begin{equation}
    \frac{\mathrm{Tr}(\hat{S}_x^2\hat{\rho}_{\text{ss}}^{\text{II}})}{(N/2)^{-1}}=\tilde{\mu}_0\braket{\hat{y}^2}_{\mu_0}+\sum_{k=1}^N\tilde{\mu}_k\braket{\hat{y}^2}_{\mu_k}.
\end{equation}
In this case the sum $\sim\sum_{k} 1$ contributes a term that scales like $N$ so it must be estimated using semiclassical analysis. However, this estimation must be done without invoking the Holstein-Primakoff approximation because the rest of eigenstates of $\hat{\rho}_{\text{ss}}^{\text{II}}$ are not necessarily polarized along $y$, as is the case for $\ket{\mu_0}$. By definition,
\begin{equation}\label{eqn:CRFBulkSum}
    \mathrm{Tr}(\hat{\rho}_{\text{ss}}^{\text{II}}\hat{S}_x^2)=\mu_0\langle\mu_0|\hat{S}_x^2|\mu_0\rangle+\sum_k \mu_k\langle\mu_k|\hat{S}_x^2|\mu_k\rangle.
\end{equation}
Using Holstein-Primakoff for $\ket{\mu_0}$ and semiclassical methods for the other $\ket{\mu_k}$ leads to (see Appendix \ref{sec:AppBulkCRf} for the derivation or Appendix \ref{sec:AppContrastVariance} for another expression valid down to $\eta=0$).
\begin{equation}\label{eqn:CRFSxVariance}
    \braket{\hat{S}_x^2}\approx \bigg(\frac{N}{4}\bigg)^{2/3}\sqrt{|\eta|}\bigg(1+\frac{2N}{3\pi}e^{-(8/3)|\eta|^{3/2}}\bigg).
\end{equation}
\begin{figure}
    \centering
    \includegraphics[width=0.48\textwidth]{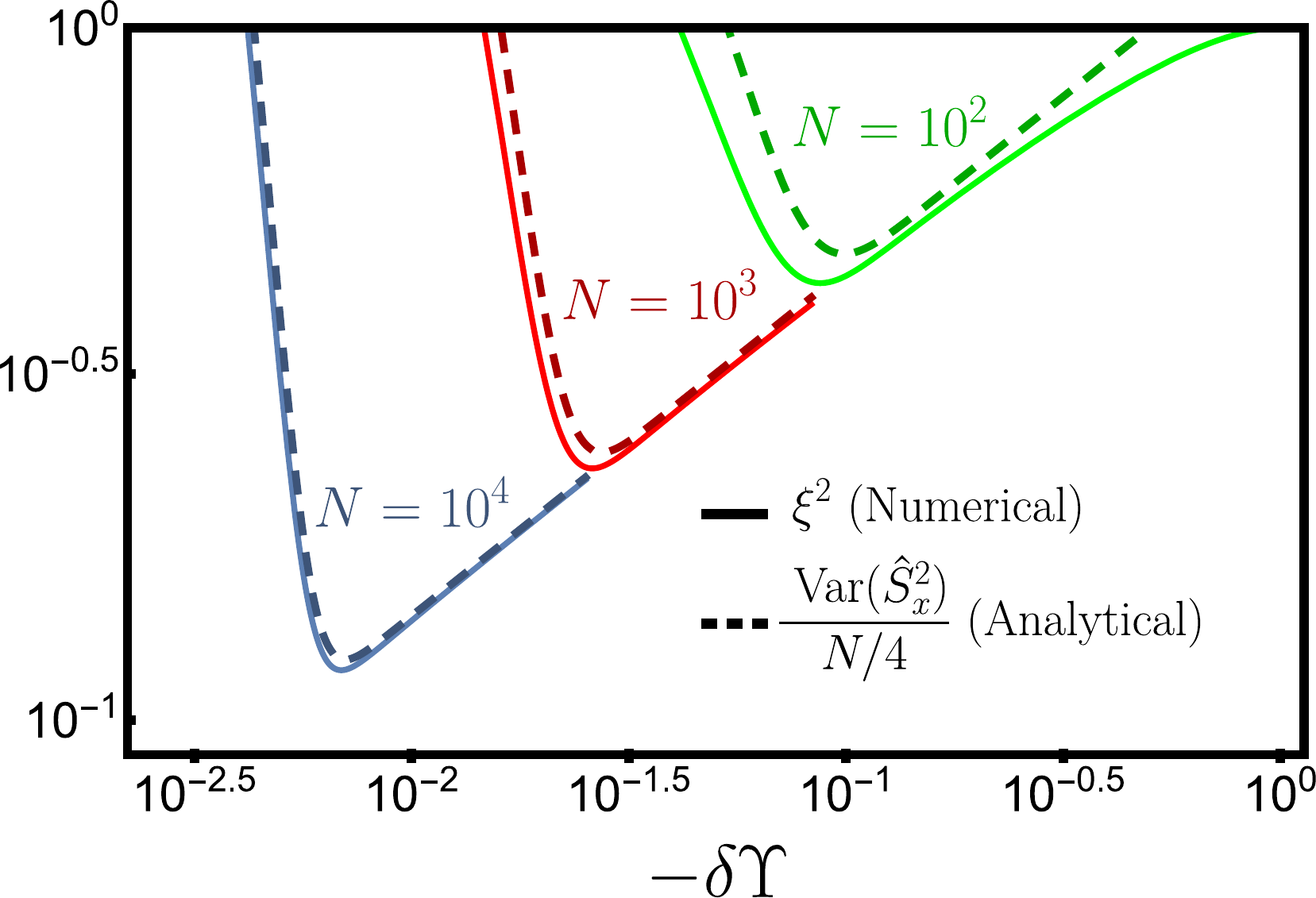}
    \caption{Squeezing $\xi^2$, calculated numerically from Eq.~(\ref{eqn:CRFSteadyState}), compared to the variance of $\hat{S}_x$ given by Eq.~(\ref{eqn:CRFSxVariance}), as a function of the distance (from below) to the critical point $-\delta\Upsilon=1-\Upsilon$ for different values of $N=10^2,10^3,10^4$. As $N$ increases, the analytical formula becomes a better estimator of the squeezing close to its optimal value.}
    \label{fig:CRFSqueezing}
\end{figure}
To compute spin squeezing we also need the contrast, which is obtained from $\braket{\hat{S}_z}$ and $\braket{\hat{S}_y}$ (see Appendix~\ref{sec:AppContrast})
\begin{align}\label{eqn:CRFContrast}
    \begin{split}
        \braket{\hat{S}_z}&=\bigg(\frac{N}{4}\bigg)^{2/3}\bigg(\frac{\int_0^\infty e^{-v^6/6-2\eta v^2}v^2\,dv}{\int_0^\infty e^{-v^6/6-2\eta v^2}\,dv}\bigg)\\
        \braket{\hat{S}_y}-\frac{N}{2}&=\frac{(2N)^{1/3}\eta}{2}-\sqrt{\frac{2}{\pi}}\frac{(N/4)^{2/3}}{\int_0^\infty e^{-v^6/6-2\eta v^2}\,dv}
    \end{split}
\end{align}
While $\hat{S}_z$ shows a consistent $N^{2/3}$ scaling as $\eta$ approaches 0, the behaviour of $\braket{\hat{S}_y}-N/2$ resembles more $\braket{\hat{S}_x^2}$: there is a very fast switch in scaling from $N^{1/3}$ at larger $\eta$ to $N^{2/3}$ at smaller $\eta$. In any case, the Bloch vector remains polarized with length $N/2$ and leading corrections are, at most, of size $N^{2/3}$. As a consequence, the squeezing parameter is determined entirely by the variance of $\hat{S}_x$
\begin{equation}
    \xi^2=\frac{N\text{Var}(\hat{S}_x)}{\braket{\hat{S}_z}^2+\braket{\hat{S}_y}^2}\approx \frac{\text{Var}(\hat{S}_x)}{N/4}.
\end{equation}
This is shown in Fig.~\ref{fig:CRFSqueezing}, where $\xi^2$ is calculated numerically and compared against the variance given by Eq.~(\ref{eqn:CRFSxVariance}) for $N=10^2,10^3,10^4$. We see that the minimum of $\xi^2$ is determined entirely by the minimum in $\hat{S}_x^2$. We thus proceed to directly minimize Eq.~(\ref{eqn:CRFSxVariance}) with respect to $\eta$. This leads to
\begin{align}\begin{split}\label{eqn:CRFMinimumSqueezingeta}
    |\eta|_{\text{min}}&=\Bigg[\frac{1}{8}-\frac{3}{8}W_{-1}\bigg(-\frac{\pi e^{1/3}}{2N}\bigg)\Bigg]^{2/3}\\
    |\eta|_{\text{min}}&\approx \bigg[\frac{3}{8}\log\bigg(\frac{2N}{\pi}\bigg)\bigg]^{2/3}\bigg[1+O\bigg(\frac{\log\log N}{\log N}\bigg)\bigg].
\end{split}\end{align}
As $N$ increases, so does $|\eta|_{\text{min}}$ (albeit very slowly) thus guaranteeing the assumption $\eta\lesssim -1$ under which all these expressions were derived. The associated minimum squeezing is
\begin{align}\label{eqn:CRFMinimumSqueezing}
    \begin{split}
        \xi^2_{\text{min}}&=\bigg(\frac{N}{4}\bigg)^{-1/3}\sqrt{|\eta|_{\text{min}}}\bigg(\frac{8|\eta|_{\text{min}}^{3/2}}{8|\eta|_{\text{min}}^{3/2}-1}\bigg)\\[3pt]
        \xi^2_{\text{min}}&\approx \Bigg[\frac{3}{2N}\log\bigg(\frac{2N}{\pi}\bigg)\Bigg]^{1/3}.
    \end{split}
\end{align}
The optimal squeezing has a prefactor $N^{-1/3}$, which was already reported in~\cite{Somech2022}, and arises in our framework from the scalings in Eq.~(\ref{eqn:Rescaling}). However, as in SSR, there is a logarithmic correction that is hard to observe numerically for any reasonable value of $N$. The first lines of Eqs.~(\ref{eqn:CRFMinimumSqueezingeta}) and ~(\ref{eqn:CRFMinimumSqueezing}) are more accurate for any realistic particle number, with corrections to those expressions being of relative size $N^{-1/3}$ instead. These corrections can still be big for moderate $N$. However, the trend towards better agreement is clear in Fig.~\ref{fig:CRFSqueezing}.\\
\section{Conclusions}
We have shown how to calculate analytically the finite-size behaviour of two all-to-all models close to their steady state phase transition points by using the Holstein-Primakoff representation of spin operators and keeping the relevant non-linearities. This allowed us to describe the rapid switch from a pure steady state to a mixed steady state in a very narrow parameter region near the phase transition points. Optimization of quantities like spin squeezing in the presence of this phenomenon gave rise to non-power-law finite size corrections that we were able to characterize theoretically.

In a realistic setting, reaching the steady state does not require fine tuned initial state preparation because the system approaches the steady state naturally. However, the associated relaxation time can be large, especially near phase transition points, where our results apply. Fortunately, in cavity QED implementations, where these models can be naturally engineered, this can be ameliorated by looking at the light leaking out of the cavity.

\section{Acknowledgements }
We thank E. Y. Song and J. T. Young for a careful reading and comments on the manuscript. We acknowledge support from the Air Force Office of Scientific Research Grants No. FA9550-18-1-0319 and No. FA9550-19-1-0275; from the Vannevar Bush Faculty Fellowship; from the Defense Advanced Research Projects Agency and Army Research Office Grant No. W911NF-16-1-0576; from the National Science Foundation JILA-Physics Frontier Center Grant No. PHY-2317149; from Quantum Leap Challenge Institute - Office of Multidisciplinary Activities Grant No. 2016244; from the US Department of Energy, Office of Science, NQI Science Research Centers, Quantum Systems Accelerator (QSA); and from the National Institute of Standards and Technology.

\bibliography{library}

\begin{thebibliography}{68}%
\makeatletter
\providecommand \@ifxundefined [1]{%
 \@ifx{#1\undefined}
}%
\providecommand \@ifnum [1]{%
 \ifnum #1\expandafter \@firstoftwo
 \else \expandafter \@secondoftwo
 \fi
}%
\providecommand \@ifx [1]{%
 \ifx #1\expandafter \@firstoftwo
 \else \expandafter \@secondoftwo
 \fi
}%
\providecommand \natexlab [1]{#1}%
\providecommand \enquote  [1]{``#1''}%
\providecommand \bibnamefont  [1]{#1}%
\providecommand \bibfnamefont [1]{#1}%
\providecommand \citenamefont [1]{#1}%
\providecommand \href@noop [0]{\@secondoftwo}%
\providecommand \href [0]{\begingroup \@sanitize@url \@href}%
\providecommand \@href[1]{\@@startlink{#1}\@@href}%
\providecommand \@@href[1]{\endgroup#1\@@endlink}%
\providecommand \@sanitize@url [0]{\catcode `\\12\catcode `\$12\catcode
  `\&12\catcode `\#12\catcode `\^12\catcode `\_12\catcode `\%12\relax}%
\providecommand \@@startlink[1]{}%
\providecommand \@@endlink[0]{}%
\providecommand \url  [0]{\begingroup\@sanitize@url \@url }%
\providecommand \@url [1]{\endgroup\@href {#1}{\urlprefix }}%
\providecommand \urlprefix  [0]{URL }%
\providecommand \Eprint [0]{\href }%
\providecommand \doibase [0]{https://doi.org/}%
\providecommand \selectlanguage [0]{\@gobble}%
\providecommand \bibinfo  [0]{\@secondoftwo}%
\providecommand \bibfield  [0]{\@secondoftwo}%
\providecommand \translation [1]{[#1]}%
\providecommand \BibitemOpen [0]{}%
\providecommand \bibitemStop [0]{}%
\providecommand \bibitemNoStop [0]{.\EOS\space}%
\providecommand \EOS [0]{\spacefactor3000\relax}%
\providecommand \BibitemShut  [1]{\csname bibitem#1\endcsname}%
\let\auto@bib@innerbib\@empty
\bibitem [{\citenamefont {Breuer}\ \emph {et~al.}(2002)\citenamefont {Breuer},
  \citenamefont {Petruccione},\ and\ \citenamefont
  {Petruccione}}]{breuer2002theory}%
  \BibitemOpen
  \bibfield  {author} {\bibinfo {author} {\bibfnamefont {H.}~\bibnamefont
  {Breuer}}, \bibinfo {author} {\bibfnamefont {F.}~\bibnamefont
  {Petruccione}},\ and\ \bibinfo {author} {\bibfnamefont {S.}~\bibnamefont
  {Petruccione}},\ }\href {https://books.google.com/books?id=0Yx5VzaMYm8C}
  {\emph {\bibinfo {title} {The Theory of Open Quantum Systems}}}\ (\bibinfo
  {publisher} {Oxford University Press},\ \bibinfo {year} {2002})\BibitemShut
  {NoStop}%
\bibitem [{\citenamefont {Degen}\ \emph {et~al.}(2017)\citenamefont {Degen},
  \citenamefont {Reinhard},\ and\ \citenamefont {Cappellaro}}]{Cappellaro2017}%
  \BibitemOpen
  \bibfield  {author} {\bibinfo {author} {\bibfnamefont {C.~L.}\ \bibnamefont
  {Degen}}, \bibinfo {author} {\bibfnamefont {F.}~\bibnamefont {Reinhard}},\
  and\ \bibinfo {author} {\bibfnamefont {P.}~\bibnamefont {Cappellaro}},\
  }\bibfield  {title} {\bibinfo {title} {Quantum sensing},\ }\href
  {https://doi.org/10.1103/RevModPhys.89.035002} {\bibfield  {journal}
  {\bibinfo  {journal} {Rev. Mod. Phys.}\ }\textbf {\bibinfo {volume} {89}},\
  \bibinfo {pages} {035002} (\bibinfo {year} {2017})}\BibitemShut {NoStop}%
\bibitem [{\citenamefont {Altman}\ \emph {et~al.}(2021)\citenamefont {Altman},
  \citenamefont {Brown}, \citenamefont {Carleo}, \citenamefont {Carr},
  \citenamefont {Demler}, \citenamefont {Chin}, \citenamefont {DeMarco},
  \citenamefont {Economou}, \citenamefont {Eriksson}, \citenamefont {Fu},
  \citenamefont {Greiner}, \citenamefont {Hazzard}, \citenamefont {Hulet},
  \citenamefont {Koll\'ar}, \citenamefont {Lev}, \citenamefont {Lukin},
  \citenamefont {Ma}, \citenamefont {Mi}, \citenamefont {Misra}, \citenamefont
  {Monroe}, \citenamefont {Murch}, \citenamefont {Nazario}, \citenamefont {Ni},
  \citenamefont {Potter}, \citenamefont {Roushan}, \citenamefont {Saffman},
  \citenamefont {Schleier-Smith}, \citenamefont {Siddiqi}, \citenamefont
  {Simmonds}, \citenamefont {Singh}, \citenamefont {Spielman}, \citenamefont
  {Temme}, \citenamefont {Weiss}, \citenamefont {Vu\ifmmode \check{c}\else
  \v{c}\fi{}kovi\ifmmode~\acute{c}\else \'{c}\fi{}}, \citenamefont
  {Vuleti\ifmmode~\acute{c}\else \'{c}\fi{}}, \citenamefont {Ye},\ and\
  \citenamefont {Zwierlein}}]{Altman2021}%
  \BibitemOpen
  \bibfield  {author} {\bibinfo {author} {\bibfnamefont {E.}~\bibnamefont
  {Altman}}, \bibinfo {author} {\bibfnamefont {K.~R.}\ \bibnamefont {Brown}},
  \bibinfo {author} {\bibfnamefont {G.}~\bibnamefont {Carleo}}, \bibinfo
  {author} {\bibfnamefont {L.~D.}\ \bibnamefont {Carr}}, \bibinfo {author}
  {\bibfnamefont {E.}~\bibnamefont {Demler}}, \bibinfo {author} {\bibfnamefont
  {C.}~\bibnamefont {Chin}}, \bibinfo {author} {\bibfnamefont {B.}~\bibnamefont
  {DeMarco}}, \bibinfo {author} {\bibfnamefont {S.~E.}\ \bibnamefont
  {Economou}}, \bibinfo {author} {\bibfnamefont {M.~A.}\ \bibnamefont
  {Eriksson}}, \bibinfo {author} {\bibfnamefont {K.-M.~C.}\ \bibnamefont {Fu}},
  \bibinfo {author} {\bibfnamefont {M.}~\bibnamefont {Greiner}}, \bibinfo
  {author} {\bibfnamefont {K.~R.}\ \bibnamefont {Hazzard}}, \bibinfo {author}
  {\bibfnamefont {R.~G.}\ \bibnamefont {Hulet}}, \bibinfo {author}
  {\bibfnamefont {A.~J.}\ \bibnamefont {Koll\'ar}}, \bibinfo {author}
  {\bibfnamefont {B.~L.}\ \bibnamefont {Lev}}, \bibinfo {author} {\bibfnamefont
  {M.~D.}\ \bibnamefont {Lukin}}, \bibinfo {author} {\bibfnamefont
  {R.}~\bibnamefont {Ma}}, \bibinfo {author} {\bibfnamefont {X.}~\bibnamefont
  {Mi}}, \bibinfo {author} {\bibfnamefont {S.}~\bibnamefont {Misra}}, \bibinfo
  {author} {\bibfnamefont {C.}~\bibnamefont {Monroe}}, \bibinfo {author}
  {\bibfnamefont {K.}~\bibnamefont {Murch}}, \bibinfo {author} {\bibfnamefont
  {Z.}~\bibnamefont {Nazario}}, \bibinfo {author} {\bibfnamefont {K.-K.}\
  \bibnamefont {Ni}}, \bibinfo {author} {\bibfnamefont {A.~C.}\ \bibnamefont
  {Potter}}, \bibinfo {author} {\bibfnamefont {P.}~\bibnamefont {Roushan}},
  \bibinfo {author} {\bibfnamefont {M.}~\bibnamefont {Saffman}}, \bibinfo
  {author} {\bibfnamefont {M.}~\bibnamefont {Schleier-Smith}}, \bibinfo
  {author} {\bibfnamefont {I.}~\bibnamefont {Siddiqi}}, \bibinfo {author}
  {\bibfnamefont {R.}~\bibnamefont {Simmonds}}, \bibinfo {author}
  {\bibfnamefont {M.}~\bibnamefont {Singh}}, \bibinfo {author} {\bibfnamefont
  {I.}~\bibnamefont {Spielman}}, \bibinfo {author} {\bibfnamefont
  {K.}~\bibnamefont {Temme}}, \bibinfo {author} {\bibfnamefont {D.~S.}\
  \bibnamefont {Weiss}}, \bibinfo {author} {\bibfnamefont {J.}~\bibnamefont
  {Vu\ifmmode \check{c}\else \v{c}\fi{}kovi\ifmmode~\acute{c}\else
  \'{c}\fi{}}}, \bibinfo {author} {\bibfnamefont {V.}~\bibnamefont
  {Vuleti\ifmmode~\acute{c}\else \'{c}\fi{}}}, \bibinfo {author} {\bibfnamefont
  {J.}~\bibnamefont {Ye}},\ and\ \bibinfo {author} {\bibfnamefont
  {M.}~\bibnamefont {Zwierlein}},\ }\bibfield  {title} {\bibinfo {title}
  {Quantum simulators: Architectures and opportunities},\ }\href
  {https://doi.org/10.1103/PRXQuantum.2.017003} {\bibfield  {journal} {\bibinfo
   {journal} {PRX Quantum}\ }\textbf {\bibinfo {volume} {2}},\ \bibinfo {pages}
  {017003} (\bibinfo {year} {2021})}\BibitemShut {NoStop}%
\bibitem [{\citenamefont {Georgescu}\ \emph {et~al.}(2014)\citenamefont
  {Georgescu}, \citenamefont {Ashhab},\ and\ \citenamefont
  {Nori}}]{Georgescu2014}%
  \BibitemOpen
  \bibfield  {author} {\bibinfo {author} {\bibfnamefont {I.~M.}\ \bibnamefont
  {Georgescu}}, \bibinfo {author} {\bibfnamefont {S.}~\bibnamefont {Ashhab}},\
  and\ \bibinfo {author} {\bibfnamefont {F.}~\bibnamefont {Nori}},\ }\bibfield
  {title} {\bibinfo {title} {Quantum simulation},\ }\href
  {https://doi.org/10.1103/RevModPhys.86.153} {\bibfield  {journal} {\bibinfo
  {journal} {Rev. Mod. Phys.}\ }\textbf {\bibinfo {volume} {86}},\ \bibinfo
  {pages} {153} (\bibinfo {year} {2014})}\BibitemShut {NoStop}%
\bibitem [{\citenamefont {Bharti}\ \emph {et~al.}(2022)\citenamefont {Bharti},
  \citenamefont {Cervera-Lierta}, \citenamefont {Kyaw}, \citenamefont {Haug},
  \citenamefont {Alperin-Lea}, \citenamefont {Anand}, \citenamefont {Degroote},
  \citenamefont {Heimonen}, \citenamefont {Kottmann}, \citenamefont {Menke},
  \citenamefont {Mok}, \citenamefont {Sim}, \citenamefont {Kwek},\ and\
  \citenamefont {Aspuru-Guzik}}]{Bharti2022}%
  \BibitemOpen
  \bibfield  {author} {\bibinfo {author} {\bibfnamefont {K.}~\bibnamefont
  {Bharti}}, \bibinfo {author} {\bibfnamefont {A.}~\bibnamefont
  {Cervera-Lierta}}, \bibinfo {author} {\bibfnamefont {T.~H.}\ \bibnamefont
  {Kyaw}}, \bibinfo {author} {\bibfnamefont {T.}~\bibnamefont {Haug}}, \bibinfo
  {author} {\bibfnamefont {S.}~\bibnamefont {Alperin-Lea}}, \bibinfo {author}
  {\bibfnamefont {A.}~\bibnamefont {Anand}}, \bibinfo {author} {\bibfnamefont
  {M.}~\bibnamefont {Degroote}}, \bibinfo {author} {\bibfnamefont
  {H.}~\bibnamefont {Heimonen}}, \bibinfo {author} {\bibfnamefont {J.~S.}\
  \bibnamefont {Kottmann}}, \bibinfo {author} {\bibfnamefont {T.}~\bibnamefont
  {Menke}}, \bibinfo {author} {\bibfnamefont {W.-K.}\ \bibnamefont {Mok}},
  \bibinfo {author} {\bibfnamefont {S.}~\bibnamefont {Sim}}, \bibinfo {author}
  {\bibfnamefont {L.-C.}\ \bibnamefont {Kwek}},\ and\ \bibinfo {author}
  {\bibfnamefont {A.}~\bibnamefont {Aspuru-Guzik}},\ }\bibfield  {title}
  {\bibinfo {title} {Noisy intermediate-scale quantum algorithms},\ }\href
  {https://doi.org/10.1103/RevModPhys.94.015004} {\bibfield  {journal}
  {\bibinfo  {journal} {Rev. Mod. Phys.}\ }\textbf {\bibinfo {volume} {94}},\
  \bibinfo {pages} {015004} (\bibinfo {year} {2022})}\BibitemShut {NoStop}%
\bibitem [{\citenamefont {de~Vega}\ and\ \citenamefont
  {Alonso}(2017)}]{Vega2017}%
  \BibitemOpen
  \bibfield  {author} {\bibinfo {author} {\bibfnamefont {I.}~\bibnamefont
  {de~Vega}}\ and\ \bibinfo {author} {\bibfnamefont {D.}~\bibnamefont
  {Alonso}},\ }\bibfield  {title} {\bibinfo {title} {Dynamics of non-markovian
  open quantum systems},\ }\href {https://doi.org/10.1103/RevModPhys.89.015001}
  {\bibfield  {journal} {\bibinfo  {journal} {Rev. Mod. Phys.}\ }\textbf
  {\bibinfo {volume} {89}},\ \bibinfo {pages} {015001} (\bibinfo {year}
  {2017})}\BibitemShut {NoStop}%
\bibitem [{\citenamefont {Helmrich}\ \emph {et~al.}(2020)\citenamefont
  {Helmrich}, \citenamefont {Arias}, \citenamefont {Lochead}, \citenamefont
  {Wintermantel}, \citenamefont {Buchhold}, \citenamefont {Diehl},\ and\
  \citenamefont {Whitlock}}]{Helmrich2020}%
  \BibitemOpen
  \bibfield  {author} {\bibinfo {author} {\bibfnamefont {S.}~\bibnamefont
  {Helmrich}}, \bibinfo {author} {\bibfnamefont {A.}~\bibnamefont {Arias}},
  \bibinfo {author} {\bibfnamefont {G.}~\bibnamefont {Lochead}}, \bibinfo
  {author} {\bibfnamefont {T.~M.}\ \bibnamefont {Wintermantel}}, \bibinfo
  {author} {\bibfnamefont {M.}~\bibnamefont {Buchhold}}, \bibinfo {author}
  {\bibfnamefont {S.}~\bibnamefont {Diehl}},\ and\ \bibinfo {author}
  {\bibfnamefont {S.}~\bibnamefont {Whitlock}},\ }\bibfield  {title} {\bibinfo
  {title} {Signatures of self-organized criticality in an ultracold atomic
  gas},\ }\href {https://doi.org/10.1038/s41586-019-1908-6} {\bibfield
  {journal} {\bibinfo  {journal} {Nature}\ }\textbf {\bibinfo {volume} {577}},\
  \bibinfo {pages} {481} (\bibinfo {year} {2020})}\BibitemShut {NoStop}%
\bibitem [{\citenamefont {Bohnet}\ \emph {et~al.}(2012)\citenamefont {Bohnet},
  \citenamefont {Chen}, \citenamefont {Weiner}, \citenamefont {Meiser},
  \citenamefont {Holland},\ and\ \citenamefont {Thompson}}]{Bohnet2012}%
  \BibitemOpen
  \bibfield  {author} {\bibinfo {author} {\bibfnamefont {J.~G.}\ \bibnamefont
  {Bohnet}}, \bibinfo {author} {\bibfnamefont {Z.}~\bibnamefont {Chen}},
  \bibinfo {author} {\bibfnamefont {J.~M.}\ \bibnamefont {Weiner}}, \bibinfo
  {author} {\bibfnamefont {D.}~\bibnamefont {Meiser}}, \bibinfo {author}
  {\bibfnamefont {M.~J.}\ \bibnamefont {Holland}},\ and\ \bibinfo {author}
  {\bibfnamefont {J.~K.}\ \bibnamefont {Thompson}},\ }\bibfield  {title}
  {\bibinfo {title} {A steady-state superradiant laser with less than one
  intracavity photon},\ }\href {https://doi.org/10.1038/nature10920} {\bibfield
   {journal} {\bibinfo  {journal} {Nature}\ }\textbf {\bibinfo {volume}
  {484}},\ \bibinfo {pages} {78} (\bibinfo {year} {2012})}\BibitemShut
  {NoStop}%
\bibitem [{\citenamefont {Meiser}\ \emph {et~al.}(2009)\citenamefont {Meiser},
  \citenamefont {Ye}, \citenamefont {Carlson},\ and\ \citenamefont
  {Holland}}]{Meiser2009}%
  \BibitemOpen
  \bibfield  {author} {\bibinfo {author} {\bibfnamefont {D.}~\bibnamefont
  {Meiser}}, \bibinfo {author} {\bibfnamefont {J.}~\bibnamefont {Ye}}, \bibinfo
  {author} {\bibfnamefont {D.~R.}\ \bibnamefont {Carlson}},\ and\ \bibinfo
  {author} {\bibfnamefont {M.~J.}\ \bibnamefont {Holland}},\ }\bibfield
  {title} {\bibinfo {title} {Prospects for a millihertz-linewidth laser},\
  }\href {https://doi.org/10.1103/PhysRevLett.102.163601} {\bibfield  {journal}
  {\bibinfo  {journal} {Phys. Rev. Lett.}\ }\textbf {\bibinfo {volume} {102}},\
  \bibinfo {pages} {163601} (\bibinfo {year} {2009})}\BibitemShut {NoStop}%
\bibitem [{\citenamefont {Gong}\ \emph {et~al.}(2018)\citenamefont {Gong},
  \citenamefont {Hamazaki},\ and\ \citenamefont {Ueda}}]{Gong2018}%
  \BibitemOpen
  \bibfield  {author} {\bibinfo {author} {\bibfnamefont {Z.}~\bibnamefont
  {Gong}}, \bibinfo {author} {\bibfnamefont {R.}~\bibnamefont {Hamazaki}},\
  and\ \bibinfo {author} {\bibfnamefont {M.}~\bibnamefont {Ueda}},\ }\bibfield
  {title} {\bibinfo {title} {Discrete time-crystalline order in cavity and
  circuit qed systems},\ }\href
  {https://doi.org/10.1103/PhysRevLett.120.040404} {\bibfield  {journal}
  {\bibinfo  {journal} {Phys. Rev. Lett.}\ }\textbf {\bibinfo {volume} {120}},\
  \bibinfo {pages} {040404} (\bibinfo {year} {2018})}\BibitemShut {NoStop}%
\bibitem [{\citenamefont {Iemini}\ \emph {et~al.}(2018)\citenamefont {Iemini},
  \citenamefont {Russomanno}, \citenamefont {Keeling}, \citenamefont
  {Schir\`o}, \citenamefont {Dalmonte},\ and\ \citenamefont
  {Fazio}}]{Iemini2018}%
  \BibitemOpen
  \bibfield  {author} {\bibinfo {author} {\bibfnamefont {F.}~\bibnamefont
  {Iemini}}, \bibinfo {author} {\bibfnamefont {A.}~\bibnamefont {Russomanno}},
  \bibinfo {author} {\bibfnamefont {J.}~\bibnamefont {Keeling}}, \bibinfo
  {author} {\bibfnamefont {M.}~\bibnamefont {Schir\`o}}, \bibinfo {author}
  {\bibfnamefont {M.}~\bibnamefont {Dalmonte}},\ and\ \bibinfo {author}
  {\bibfnamefont {R.}~\bibnamefont {Fazio}},\ }\bibfield  {title} {\bibinfo
  {title} {Boundary time crystals},\ }\href
  {https://doi.org/10.1103/PhysRevLett.121.035301} {\bibfield  {journal}
  {\bibinfo  {journal} {Phys. Rev. Lett.}\ }\textbf {\bibinfo {volume} {121}},\
  \bibinfo {pages} {035301} (\bibinfo {year} {2018})}\BibitemShut {NoStop}%
\bibitem [{\citenamefont {Zhu}\ \emph {et~al.}(2015)\citenamefont {Zhu},
  \citenamefont {Schachenmayer}, \citenamefont {Xu}, \citenamefont {Herrera},
  \citenamefont {Restrepo}, \citenamefont {Holland},\ and\ \citenamefont
  {Rey}}]{Zhu_2015}%
  \BibitemOpen
  \bibfield  {author} {\bibinfo {author} {\bibfnamefont {B.}~\bibnamefont
  {Zhu}}, \bibinfo {author} {\bibfnamefont {J.}~\bibnamefont {Schachenmayer}},
  \bibinfo {author} {\bibfnamefont {M.}~\bibnamefont {Xu}}, \bibinfo {author}
  {\bibfnamefont {F.}~\bibnamefont {Herrera}}, \bibinfo {author} {\bibfnamefont
  {J.~G.}\ \bibnamefont {Restrepo}}, \bibinfo {author} {\bibfnamefont {M.~J.}\
  \bibnamefont {Holland}},\ and\ \bibinfo {author} {\bibfnamefont {A.~M.}\
  \bibnamefont {Rey}},\ }\bibfield  {title} {\bibinfo {title} {Synchronization
  of interacting quantum dipoles},\ }\href
  {https://doi.org/10.1088/1367-2630/17/8/083063} {\bibfield  {journal}
  {\bibinfo  {journal} {New Journal of Physics}\ }\textbf {\bibinfo {volume}
  {17}},\ \bibinfo {pages} {083063} (\bibinfo {year} {2015})}\BibitemShut
  {NoStop}%
\bibitem [{\citenamefont {Owen}\ \emph {et~al.}(2018)\citenamefont {Owen},
  \citenamefont {Jin}, \citenamefont {Rossini}, \citenamefont {Fazio},\ and\
  \citenamefont {Hartmann}}]{Owen2018}%
  \BibitemOpen
  \bibfield  {author} {\bibinfo {author} {\bibfnamefont {E.~T.}\ \bibnamefont
  {Owen}}, \bibinfo {author} {\bibfnamefont {J.}~\bibnamefont {Jin}}, \bibinfo
  {author} {\bibfnamefont {D.}~\bibnamefont {Rossini}}, \bibinfo {author}
  {\bibfnamefont {R.}~\bibnamefont {Fazio}},\ and\ \bibinfo {author}
  {\bibfnamefont {M.~J.}\ \bibnamefont {Hartmann}},\ }\bibfield  {title}
  {\bibinfo {title} {Quantum correlations and limit cycles in the
  driven-dissipative heisenberg lattice},\ }\href
  {https://doi.org/10.1088/1367-2630/aab7d3} {\bibfield  {journal} {\bibinfo
  {journal} {New Journal of Physics}\ }\textbf {\bibinfo {volume} {20}},\
  \bibinfo {pages} {045004} (\bibinfo {year} {2018})}\BibitemShut {NoStop}%
\bibitem [{\citenamefont {Link}\ and\ \citenamefont {Strunz}(2020)}]{Link2020}%
  \BibitemOpen
  \bibfield  {author} {\bibinfo {author} {\bibfnamefont {V.}~\bibnamefont
  {Link}}\ and\ \bibinfo {author} {\bibfnamefont {W.~T.}\ \bibnamefont
  {Strunz}},\ }\bibfield  {title} {\bibinfo {title} {Dynamical phase
  transitions in dissipative quantum dynamics with quantum optical
  realization},\ }\href {https://doi.org/10.1103/PhysRevLett.125.143602}
  {\bibfield  {journal} {\bibinfo  {journal} {Phys. Rev. Lett.}\ }\textbf
  {\bibinfo {volume} {125}},\ \bibinfo {pages} {143602} (\bibinfo {year}
  {2020})}\BibitemShut {NoStop}%
\bibitem [{\citenamefont {Eisert}\ and\ \citenamefont
  {Prosen}(2010)}]{Eisert2010}%
  \BibitemOpen
  \bibfield  {author} {\bibinfo {author} {\bibfnamefont {J.}~\bibnamefont
  {Eisert}}\ and\ \bibinfo {author} {\bibfnamefont {T.}~\bibnamefont
  {Prosen}},\ }\href {https://doi.org/10.48550/ARXIV.1012.5013} {\bibinfo
  {title} {Noise-driven quantum criticality}} (\bibinfo {year}
  {2010})\BibitemShut {NoStop}%
\bibitem [{\citenamefont {Maghrebi}\ and\ \citenamefont
  {Gorshkov}(2016)}]{Maghrebi2016}%
  \BibitemOpen
  \bibfield  {author} {\bibinfo {author} {\bibfnamefont {M.~F.}\ \bibnamefont
  {Maghrebi}}\ and\ \bibinfo {author} {\bibfnamefont {A.~V.}\ \bibnamefont
  {Gorshkov}},\ }\bibfield  {title} {\bibinfo {title} {Nonequilibrium many-body
  steady states via keldysh formalism},\ }\href
  {https://doi.org/10.1103/PhysRevB.93.014307} {\bibfield  {journal} {\bibinfo
  {journal} {Phys. Rev. B}\ }\textbf {\bibinfo {volume} {93}},\ \bibinfo
  {pages} {014307} (\bibinfo {year} {2016})}\BibitemShut {NoStop}%
\bibitem [{\citenamefont {Young}\ \emph {et~al.}(2020)\citenamefont {Young},
  \citenamefont {Gorshkov}, \citenamefont {Foss-Feig},\ and\ \citenamefont
  {Maghrebi}}]{Young2020}%
  \BibitemOpen
  \bibfield  {author} {\bibinfo {author} {\bibfnamefont {J.~T.}\ \bibnamefont
  {Young}}, \bibinfo {author} {\bibfnamefont {A.~V.}\ \bibnamefont {Gorshkov}},
  \bibinfo {author} {\bibfnamefont {M.}~\bibnamefont {Foss-Feig}},\ and\
  \bibinfo {author} {\bibfnamefont {M.~F.}\ \bibnamefont {Maghrebi}},\
  }\bibfield  {title} {\bibinfo {title} {Nonequilibrium fixed points of coupled
  ising models},\ }\href {https://doi.org/10.1103/PhysRevX.10.011039}
  {\bibfield  {journal} {\bibinfo  {journal} {Phys. Rev. X}\ }\textbf {\bibinfo
  {volume} {10}},\ \bibinfo {pages} {011039} (\bibinfo {year}
  {2020})}\BibitemShut {NoStop}%
\bibitem [{\citenamefont {Marino}\ and\ \citenamefont
  {Diehl}(2016)}]{Marino2016}%
  \BibitemOpen
  \bibfield  {author} {\bibinfo {author} {\bibfnamefont {J.}~\bibnamefont
  {Marino}}\ and\ \bibinfo {author} {\bibfnamefont {S.}~\bibnamefont {Diehl}},\
  }\bibfield  {title} {\bibinfo {title} {Driven markovian quantum
  criticality},\ }\href {https://doi.org/10.1103/PhysRevLett.116.070407}
  {\bibfield  {journal} {\bibinfo  {journal} {Phys. Rev. Lett.}\ }\textbf
  {\bibinfo {volume} {116}},\ \bibinfo {pages} {070407} (\bibinfo {year}
  {2016})}\BibitemShut {NoStop}%
\bibitem [{\citenamefont {Le~Boit\'e}\ \emph {et~al.}(2013)\citenamefont
  {Le~Boit\'e}, \citenamefont {Orso},\ and\ \citenamefont
  {Ciuti}}]{LeBoite2013}%
  \BibitemOpen
  \bibfield  {author} {\bibinfo {author} {\bibfnamefont {A.}~\bibnamefont
  {Le~Boit\'e}}, \bibinfo {author} {\bibfnamefont {G.}~\bibnamefont {Orso}},\
  and\ \bibinfo {author} {\bibfnamefont {C.}~\bibnamefont {Ciuti}},\ }\bibfield
   {title} {\bibinfo {title} {Steady-state phases and tunneling-induced
  instabilities in the driven dissipative bose-hubbard model},\ }\href
  {https://doi.org/10.1103/PhysRevLett.110.233601} {\bibfield  {journal}
  {\bibinfo  {journal} {Phys. Rev. Lett.}\ }\textbf {\bibinfo {volume} {110}},\
  \bibinfo {pages} {233601} (\bibinfo {year} {2013})}\BibitemShut {NoStop}%
\bibitem [{\citenamefont {Rota}\ \emph {et~al.}(2019)\citenamefont {Rota},
  \citenamefont {Minganti}, \citenamefont {Ciuti},\ and\ \citenamefont
  {Savona}}]{Rota2019}%
  \BibitemOpen
  \bibfield  {author} {\bibinfo {author} {\bibfnamefont {R.}~\bibnamefont
  {Rota}}, \bibinfo {author} {\bibfnamefont {F.}~\bibnamefont {Minganti}},
  \bibinfo {author} {\bibfnamefont {C.}~\bibnamefont {Ciuti}},\ and\ \bibinfo
  {author} {\bibfnamefont {V.}~\bibnamefont {Savona}},\ }\bibfield  {title}
  {\bibinfo {title} {Quantum critical regime in a quadratically driven
  nonlinear photonic lattice},\ }\href
  {https://doi.org/10.1103/PhysRevLett.122.110405} {\bibfield  {journal}
  {\bibinfo  {journal} {Phys. Rev. Lett.}\ }\textbf {\bibinfo {volume} {122}},\
  \bibinfo {pages} {110405} (\bibinfo {year} {2019})}\BibitemShut {NoStop}%
\bibitem [{\citenamefont {Sieberer}\ \emph {et~al.}(2013)\citenamefont
  {Sieberer}, \citenamefont {Huber}, \citenamefont {Altman},\ and\
  \citenamefont {Diehl}}]{Sieberer2013}%
  \BibitemOpen
  \bibfield  {author} {\bibinfo {author} {\bibfnamefont {L.~M.}\ \bibnamefont
  {Sieberer}}, \bibinfo {author} {\bibfnamefont {S.~D.}\ \bibnamefont {Huber}},
  \bibinfo {author} {\bibfnamefont {E.}~\bibnamefont {Altman}},\ and\ \bibinfo
  {author} {\bibfnamefont {S.}~\bibnamefont {Diehl}},\ }\bibfield  {title}
  {\bibinfo {title} {Dynamical critical phenomena in driven-dissipative
  systems},\ }\href {https://doi.org/10.1103/PhysRevLett.110.195301} {\bibfield
   {journal} {\bibinfo  {journal} {Phys. Rev. Lett.}\ }\textbf {\bibinfo
  {volume} {110}},\ \bibinfo {pages} {195301} (\bibinfo {year}
  {2013})}\BibitemShut {NoStop}%
\bibitem [{\citenamefont {Kraus}\ \emph {et~al.}(2008)\citenamefont {Kraus},
  \citenamefont {B\"uchler}, \citenamefont {Diehl}, \citenamefont {Kantian},
  \citenamefont {Micheli},\ and\ \citenamefont {Zoller}}]{Kraus2008}%
  \BibitemOpen
  \bibfield  {author} {\bibinfo {author} {\bibfnamefont {B.}~\bibnamefont
  {Kraus}}, \bibinfo {author} {\bibfnamefont {H.~P.}\ \bibnamefont
  {B\"uchler}}, \bibinfo {author} {\bibfnamefont {S.}~\bibnamefont {Diehl}},
  \bibinfo {author} {\bibfnamefont {A.}~\bibnamefont {Kantian}}, \bibinfo
  {author} {\bibfnamefont {A.}~\bibnamefont {Micheli}},\ and\ \bibinfo {author}
  {\bibfnamefont {P.}~\bibnamefont {Zoller}},\ }\bibfield  {title} {\bibinfo
  {title} {Preparation of entangled states by quantum markov processes},\
  }\href {https://doi.org/10.1103/PhysRevA.78.042307} {\bibfield  {journal}
  {\bibinfo  {journal} {Phys. Rev. A}\ }\textbf {\bibinfo {volume} {78}},\
  \bibinfo {pages} {042307} (\bibinfo {year} {2008})}\BibitemShut {NoStop}%
\bibitem [{\citenamefont {Diehl}\ \emph {et~al.}(2008)\citenamefont {Diehl},
  \citenamefont {Micheli}, \citenamefont {Kantian}, \citenamefont {Kraus},
  \citenamefont {B{\"u}chler},\ and\ \citenamefont {Zoller}}]{Diehl2008}%
  \BibitemOpen
  \bibfield  {author} {\bibinfo {author} {\bibfnamefont {S.}~\bibnamefont
  {Diehl}}, \bibinfo {author} {\bibfnamefont {A.}~\bibnamefont {Micheli}},
  \bibinfo {author} {\bibfnamefont {A.}~\bibnamefont {Kantian}}, \bibinfo
  {author} {\bibfnamefont {B.}~\bibnamefont {Kraus}}, \bibinfo {author}
  {\bibfnamefont {H.~P.}\ \bibnamefont {B{\"u}chler}},\ and\ \bibinfo {author}
  {\bibfnamefont {P.}~\bibnamefont {Zoller}},\ }\bibfield  {title} {\bibinfo
  {title} {Quantum states and phases in driven open quantum systems with cold
  atoms},\ }\href {https://doi.org/10.1038/nphys1073} {\bibfield  {journal}
  {\bibinfo  {journal} {Nature Physics}\ }\textbf {\bibinfo {volume} {4}},\
  \bibinfo {pages} {878} (\bibinfo {year} {2008})}\BibitemShut {NoStop}%
\bibitem [{\citenamefont {Verstraete}\ \emph {et~al.}(2009)\citenamefont
  {Verstraete}, \citenamefont {Wolf},\ and\ \citenamefont
  {Ignacio~Cirac}}]{Verstraete2009}%
  \BibitemOpen
  \bibfield  {author} {\bibinfo {author} {\bibfnamefont {F.}~\bibnamefont
  {Verstraete}}, \bibinfo {author} {\bibfnamefont {M.~M.}\ \bibnamefont
  {Wolf}},\ and\ \bibinfo {author} {\bibfnamefont {J.}~\bibnamefont
  {Ignacio~Cirac}},\ }\bibfield  {title} {\bibinfo {title} {Quantum computation
  and quantum-state engineering driven by dissipation},\ }\href
  {https://doi.org/10.1038/nphys1342} {\bibfield  {journal} {\bibinfo
  {journal} {Nature Physics}\ }\textbf {\bibinfo {volume} {5}},\ \bibinfo
  {pages} {633} (\bibinfo {year} {2009})}\BibitemShut {NoStop}%
\bibitem [{\citenamefont {Carmichael}(1980)}]{Carmichael_1980}%
  \BibitemOpen
  \bibfield  {author} {\bibinfo {author} {\bibfnamefont {H.~J.}\ \bibnamefont
  {Carmichael}},\ }\bibfield  {title} {\bibinfo {title} {Analytical and
  numerical results for the steady state in cooperative resonance
  fluorescence},\ }\href {https://doi.org/10.1088/0022-3700/13/18/009}
  {\bibfield  {journal} {\bibinfo  {journal} {Journal of Physics B: Atomic and
  Molecular Physics}\ }\textbf {\bibinfo {volume} {13}},\ \bibinfo {pages}
  {3551} (\bibinfo {year} {1980})}\BibitemShut {NoStop}%
\bibitem [{\citenamefont {Morrison}\ and\ \citenamefont
  {Parkins}(2008)}]{Morrison2008}%
  \BibitemOpen
  \bibfield  {author} {\bibinfo {author} {\bibfnamefont {S.}~\bibnamefont
  {Morrison}}\ and\ \bibinfo {author} {\bibfnamefont {A.~S.}\ \bibnamefont
  {Parkins}},\ }\bibfield  {title} {\bibinfo {title} {Dynamical quantum phase
  transitions in the dissipative lipkin-meshkov-glick model with proposed
  realization in optical cavity qed},\ }\href
  {https://doi.org/10.1103/PhysRevLett.100.040403} {\bibfield  {journal}
  {\bibinfo  {journal} {Phys. Rev. Lett.}\ }\textbf {\bibinfo {volume} {100}},\
  \bibinfo {pages} {040403} (\bibinfo {year} {2008})}\BibitemShut {NoStop}%
\bibitem [{\citenamefont {Lee}\ \emph {et~al.}(2014)\citenamefont {Lee},
  \citenamefont {Chan},\ and\ \citenamefont {Yelin}}]{Lee2014}%
  \BibitemOpen
  \bibfield  {author} {\bibinfo {author} {\bibfnamefont {T.~E.}\ \bibnamefont
  {Lee}}, \bibinfo {author} {\bibfnamefont {C.-K.}\ \bibnamefont {Chan}},\ and\
  \bibinfo {author} {\bibfnamefont {S.~F.}\ \bibnamefont {Yelin}},\ }\bibfield
  {title} {\bibinfo {title} {Dissipative phase transitions: Independent versus
  collective decay and spin squeezing},\ }\href
  {https://doi.org/10.1103/PhysRevA.90.052109} {\bibfield  {journal} {\bibinfo
  {journal} {Phys. Rev. A}\ }\textbf {\bibinfo {volume} {90}},\ \bibinfo
  {pages} {052109} (\bibinfo {year} {2014})}\BibitemShut {NoStop}%
\bibitem [{\citenamefont {Kessler}\ \emph {et~al.}(2012)\citenamefont
  {Kessler}, \citenamefont {Giedke}, \citenamefont {Imamoglu}, \citenamefont
  {Yelin}, \citenamefont {Lukin},\ and\ \citenamefont {Cirac}}]{Kessler2012}%
  \BibitemOpen
  \bibfield  {author} {\bibinfo {author} {\bibfnamefont {E.~M.}\ \bibnamefont
  {Kessler}}, \bibinfo {author} {\bibfnamefont {G.}~\bibnamefont {Giedke}},
  \bibinfo {author} {\bibfnamefont {A.}~\bibnamefont {Imamoglu}}, \bibinfo
  {author} {\bibfnamefont {S.~F.}\ \bibnamefont {Yelin}}, \bibinfo {author}
  {\bibfnamefont {M.~D.}\ \bibnamefont {Lukin}},\ and\ \bibinfo {author}
  {\bibfnamefont {J.~I.}\ \bibnamefont {Cirac}},\ }\bibfield  {title} {\bibinfo
  {title} {Dissipative phase transition in a central spin system},\ }\href
  {https://doi.org/10.1103/PhysRevA.86.012116} {\bibfield  {journal} {\bibinfo
  {journal} {Phys. Rev. A}\ }\textbf {\bibinfo {volume} {86}},\ \bibinfo
  {pages} {012116} (\bibinfo {year} {2012})}\BibitemShut {NoStop}%
\bibitem [{\citenamefont {Ferreira}\ and\ \citenamefont
  {Ribeiro}(2019)}]{Ferreira2019}%
  \BibitemOpen
  \bibfield  {author} {\bibinfo {author} {\bibfnamefont {J.~a.~S.}\
  \bibnamefont {Ferreira}}\ and\ \bibinfo {author} {\bibfnamefont
  {P.}~\bibnamefont {Ribeiro}},\ }\bibfield  {title} {\bibinfo {title}
  {Lipkin-meshkov-glick model with markovian dissipation: A description of a
  collective spin on a metallic surface},\ }\href
  {https://doi.org/10.1103/PhysRevB.100.184422} {\bibfield  {journal} {\bibinfo
   {journal} {Phys. Rev. B}\ }\textbf {\bibinfo {volume} {100}},\ \bibinfo
  {pages} {184422} (\bibinfo {year} {2019})}\BibitemShut {NoStop}%
\bibitem [{\citenamefont {Pavlov}\ \emph {et~al.}(2023)\citenamefont {Pavlov},
  \citenamefont {Porras},\ and\ \citenamefont {Ivanov}}]{Pavlov2023}%
  \BibitemOpen
  \bibfield  {author} {\bibinfo {author} {\bibfnamefont {V.~P.}\ \bibnamefont
  {Pavlov}}, \bibinfo {author} {\bibfnamefont {D.}~\bibnamefont {Porras}},\
  and\ \bibinfo {author} {\bibfnamefont {P.~A.}\ \bibnamefont {Ivanov}},\
  }\bibfield  {title} {\bibinfo {title} {Quantum metrology with critical
  driven-dissipative collective spin system},\ }\href
  {https://doi.org/10.1088/1402-4896/ace99f} {\bibfield  {journal} {\bibinfo
  {journal} {Physica Scripta}\ }\textbf {\bibinfo {volume} {98}},\ \bibinfo
  {pages} {095103} (\bibinfo {year} {2023})}\BibitemShut {NoStop}%
\bibitem [{\citenamefont {Kitagawa}\ and\ \citenamefont
  {Ueda}(1993)}]{Kitagawa1993}%
  \BibitemOpen
  \bibfield  {author} {\bibinfo {author} {\bibfnamefont {M.}~\bibnamefont
  {Kitagawa}}\ and\ \bibinfo {author} {\bibfnamefont {M.}~\bibnamefont
  {Ueda}},\ }\bibfield  {title} {\bibinfo {title} {Squeezed spin states},\
  }\href {https://doi.org/10.1103/PhysRevA.47.5138} {\bibfield  {journal}
  {\bibinfo  {journal} {Phys. Rev. A}\ }\textbf {\bibinfo {volume} {47}},\
  \bibinfo {pages} {5138} (\bibinfo {year} {1993})}\BibitemShut {NoStop}%
\bibitem [{\citenamefont {Ma}\ \emph {et~al.}(2011)\citenamefont {Ma},
  \citenamefont {Wang}, \citenamefont {Sun},\ and\ \citenamefont
  {Nori}}]{MA2011}%
  \BibitemOpen
  \bibfield  {author} {\bibinfo {author} {\bibfnamefont {J.}~\bibnamefont
  {Ma}}, \bibinfo {author} {\bibfnamefont {X.}~\bibnamefont {Wang}}, \bibinfo
  {author} {\bibfnamefont {C.}~\bibnamefont {Sun}},\ and\ \bibinfo {author}
  {\bibfnamefont {F.}~\bibnamefont {Nori}},\ }\bibfield  {title} {\bibinfo
  {title} {Quantum spin squeezing},\ }\href
  {https://doi.org/https://doi.org/10.1016/j.physrep.2011.08.003} {\bibfield
  {journal} {\bibinfo  {journal} {Physics Reports}\ }\textbf {\bibinfo {volume}
  {509}},\ \bibinfo {pages} {89} (\bibinfo {year} {2011})}\BibitemShut
  {NoStop}%
\bibitem [{\citenamefont {Robinson}\ \emph {et~al.}(2022)\citenamefont
  {Robinson}, \citenamefont {Miklos}, \citenamefont {Tso}, \citenamefont
  {Kennedy}, \citenamefont {Bothwell}, \citenamefont {Kedar}, \citenamefont
  {Thompson},\ and\ \citenamefont {Ye}}]{Robinson2022}%
  \BibitemOpen
  \bibfield  {author} {\bibinfo {author} {\bibfnamefont {J.~M.}\ \bibnamefont
  {Robinson}}, \bibinfo {author} {\bibfnamefont {M.}~\bibnamefont {Miklos}},
  \bibinfo {author} {\bibfnamefont {Y.~M.}\ \bibnamefont {Tso}}, \bibinfo
  {author} {\bibfnamefont {C.~J.}\ \bibnamefont {Kennedy}}, \bibinfo {author}
  {\bibfnamefont {T.}~\bibnamefont {Bothwell}}, \bibinfo {author}
  {\bibfnamefont {D.}~\bibnamefont {Kedar}}, \bibinfo {author} {\bibfnamefont
  {J.~K.}\ \bibnamefont {Thompson}},\ and\ \bibinfo {author} {\bibfnamefont
  {J.}~\bibnamefont {Ye}},\ }\href {https://doi.org/10.48550/ARXIV.2211.08621}
  {\bibinfo {title} {Direct comparison of two spin squeezed optical clocks
  below the quantum projection noise limit}} (\bibinfo {year}
  {2022})\BibitemShut {NoStop}%
\bibitem [{\citenamefont {Pedrozo-Pe{\~{n}}afiel}\ \emph
  {et~al.}(2020)\citenamefont {Pedrozo-Pe{\~{n}}afiel}, \citenamefont
  {Colombo}, \citenamefont {Shu}, \citenamefont {Adiyatullin}, \citenamefont
  {Li}, \citenamefont {Mendez}, \citenamefont {Braverman}, \citenamefont
  {Kawasaki}, \citenamefont {Akamatsu}, \citenamefont {Xiao},\ and\
  \citenamefont {Vuleti{\'{c}}}}]{PedrozoPenafiel2020}%
  \BibitemOpen
  \bibfield  {author} {\bibinfo {author} {\bibfnamefont {E.}~\bibnamefont
  {Pedrozo-Pe{\~{n}}afiel}}, \bibinfo {author} {\bibfnamefont {S.}~\bibnamefont
  {Colombo}}, \bibinfo {author} {\bibfnamefont {C.}~\bibnamefont {Shu}},
  \bibinfo {author} {\bibfnamefont {A.~F.}\ \bibnamefont {Adiyatullin}},
  \bibinfo {author} {\bibfnamefont {Z.}~\bibnamefont {Li}}, \bibinfo {author}
  {\bibfnamefont {E.}~\bibnamefont {Mendez}}, \bibinfo {author} {\bibfnamefont
  {B.}~\bibnamefont {Braverman}}, \bibinfo {author} {\bibfnamefont
  {A.}~\bibnamefont {Kawasaki}}, \bibinfo {author} {\bibfnamefont
  {D.}~\bibnamefont {Akamatsu}}, \bibinfo {author} {\bibfnamefont
  {Y.}~\bibnamefont {Xiao}},\ and\ \bibinfo {author} {\bibfnamefont
  {V.}~\bibnamefont {Vuleti{\'{c}}}},\ }\bibfield  {title} {\bibinfo {title}
  {Entanglement on an optical atomic-clock transition},\ }\href
  {https://doi.org/10.1038/s41586-020-3006-1} {\bibfield  {journal} {\bibinfo
  {journal} {Nature}\ }\textbf {\bibinfo {volume} {588}},\ \bibinfo {pages}
  {414} (\bibinfo {year} {2020})}\BibitemShut {NoStop}%
\bibitem [{\citenamefont {Schulte}\ \emph {et~al.}(2020)\citenamefont
  {Schulte}, \citenamefont {Lisdat}, \citenamefont {Schmidt}, \citenamefont
  {Sterr},\ and\ \citenamefont {Hammerer}}]{Schulte2020}%
  \BibitemOpen
  \bibfield  {author} {\bibinfo {author} {\bibfnamefont {M.}~\bibnamefont
  {Schulte}}, \bibinfo {author} {\bibfnamefont {C.}~\bibnamefont {Lisdat}},
  \bibinfo {author} {\bibfnamefont {P.~O.}\ \bibnamefont {Schmidt}}, \bibinfo
  {author} {\bibfnamefont {U.}~\bibnamefont {Sterr}},\ and\ \bibinfo {author}
  {\bibfnamefont {K.}~\bibnamefont {Hammerer}},\ }\bibfield  {title} {\bibinfo
  {title} {Prospects and challenges for squeezing-enhanced optical atomic
  clocks},\ }\href {https://doi.org/10.1038/s41467-020-19403-7} {\bibfield
  {journal} {\bibinfo  {journal} {Nature Communications}\ }\textbf {\bibinfo
  {volume} {11}},\ \bibinfo {pages} {5955} (\bibinfo {year}
  {2020})}\BibitemShut {NoStop}%
\bibitem [{\citenamefont {Holstein}\ and\ \citenamefont
  {Primakoff}(1940)}]{Holstein1940}%
  \BibitemOpen
  \bibfield  {author} {\bibinfo {author} {\bibfnamefont {T.}~\bibnamefont
  {Holstein}}\ and\ \bibinfo {author} {\bibfnamefont {H.}~\bibnamefont
  {Primakoff}},\ }\bibfield  {title} {\bibinfo {title} {Field dependence of the
  intrinsic domain magnetization of a ferromagnet},\ }\href
  {https://doi.org/10.1103/PhysRev.58.1098} {\bibfield  {journal} {\bibinfo
  {journal} {Phys. Rev.}\ }\textbf {\bibinfo {volume} {58}},\ \bibinfo {pages}
  {1098} (\bibinfo {year} {1940})}\BibitemShut {NoStop}%
\bibitem [{\citenamefont {Hirsch}\ \emph {et~al.}(2013)\citenamefont {Hirsch},
  \citenamefont {Castaños}, \citenamefont {López-Peña},\ and\ \citenamefont
  {Nahmad-Achar}}]{Hirsch2013}%
  \BibitemOpen
  \bibfield  {author} {\bibinfo {author} {\bibfnamefont {J.~G.}\ \bibnamefont
  {Hirsch}}, \bibinfo {author} {\bibfnamefont {O.}~\bibnamefont {Castaños}},
  \bibinfo {author} {\bibfnamefont {R.}~\bibnamefont {López-Peña}},\ and\
  \bibinfo {author} {\bibfnamefont {E.}~\bibnamefont {Nahmad-Achar}},\
  }\bibfield  {title} {\bibinfo {title} {Virtues and limitations of the
  truncated holstein–primakoff description of quantum rotors},\ }\href
  {https://doi.org/10.1088/0031-8949/87/03/038106} {\bibfield  {journal}
  {\bibinfo  {journal} {Physica Scripta}\ }\textbf {\bibinfo {volume} {87}},\
  \bibinfo {pages} {038106} (\bibinfo {year} {2013})}\BibitemShut {NoStop}%
\bibitem [{\citenamefont {Reslen}\ \emph {et~al.}(2005)\citenamefont {Reslen},
  \citenamefont {Quiroga},\ and\ \citenamefont {Johnson}}]{J.Reslen_2005}%
  \BibitemOpen
  \bibfield  {author} {\bibinfo {author} {\bibfnamefont {J.}~\bibnamefont
  {Reslen}}, \bibinfo {author} {\bibfnamefont {L.}~\bibnamefont {Quiroga}},\
  and\ \bibinfo {author} {\bibfnamefont {N.~F.}\ \bibnamefont {Johnson}},\
  }\bibfield  {title} {\bibinfo {title} {Direct equivalence between quantum
  phase transition phenomena in radiation-matter and magnetic systems: Scaling
  of entanglement},\ }\href {https://doi.org/10.1209/epl/i2004-10313-4}
  {\bibfield  {journal} {\bibinfo  {journal} {Europhysics Letters}\ }\textbf
  {\bibinfo {volume} {69}},\ \bibinfo {pages} {8} (\bibinfo {year}
  {2005})}\BibitemShut {NoStop}%
\bibitem [{\citenamefont {Bastarrachea-Magnani}\ \emph
  {et~al.}(2014)\citenamefont {Bastarrachea-Magnani}, \citenamefont
  {Castaños}, \citenamefont {Nahmad-Achar}, \citenamefont {López-Peña},\
  and\ \citenamefont {Hirsch}}]{Bastarrachea-Magnani_2014}%
  \BibitemOpen
  \bibfield  {author} {\bibinfo {author} {\bibfnamefont {M.~A.}\ \bibnamefont
  {Bastarrachea-Magnani}}, \bibinfo {author} {\bibfnamefont {O.}~\bibnamefont
  {Castaños}}, \bibinfo {author} {\bibfnamefont {E.}~\bibnamefont
  {Nahmad-Achar}}, \bibinfo {author} {\bibfnamefont {R.}~\bibnamefont
  {López-Peña}},\ and\ \bibinfo {author} {\bibfnamefont {J.~G.}\ \bibnamefont
  {Hirsch}},\ }\bibfield  {title} {\bibinfo {title} {Fidelity, susceptibility
  and critical exponents in the dicke model},\ }\href
  {https://doi.org/10.1088/1742-6596/492/1/012012} {\bibfield  {journal}
  {\bibinfo  {journal} {Journal of Physics: Conference Series}\ }\textbf
  {\bibinfo {volume} {492}},\ \bibinfo {pages} {012012} (\bibinfo {year}
  {2014})}\BibitemShut {NoStop}%
\bibitem [{\citenamefont {Nahmad-Achar}\ \emph {et~al.}(2013)\citenamefont
  {Nahmad-Achar}, \citenamefont {Castaños}, \citenamefont {López-Peña},\
  and\ \citenamefont {Hirsch}}]{Nahmad-Achar_2013}%
  \BibitemOpen
  \bibfield  {author} {\bibinfo {author} {\bibfnamefont {E.}~\bibnamefont
  {Nahmad-Achar}}, \bibinfo {author} {\bibfnamefont {O.}~\bibnamefont
  {Castaños}}, \bibinfo {author} {\bibfnamefont {R.}~\bibnamefont
  {López-Peña}},\ and\ \bibinfo {author} {\bibfnamefont {J.~G.}\ \bibnamefont
  {Hirsch}},\ }\bibfield  {title} {\bibinfo {title} {Mathematical methods in
  quantum optics: the dicke model},\ }\href
  {https://doi.org/10.1088/0031-8949/87/03/038114} {\bibfield  {journal}
  {\bibinfo  {journal} {Physica Scripta}\ }\textbf {\bibinfo {volume} {87}},\
  \bibinfo {pages} {038114} (\bibinfo {year} {2013})}\BibitemShut {NoStop}%
\bibitem [{\citenamefont {Dusuel}\ and\ \citenamefont
  {Vidal}(2004)}]{Dusuel2004}%
  \BibitemOpen
  \bibfield  {author} {\bibinfo {author} {\bibfnamefont {S.}~\bibnamefont
  {Dusuel}}\ and\ \bibinfo {author} {\bibfnamefont {J.}~\bibnamefont {Vidal}},\
  }\bibfield  {title} {\bibinfo {title} {Finite-size scaling exponents of the
  lipkin-meshkov-glick model},\ }\href
  {https://doi.org/10.1103/PhysRevLett.93.237204} {\bibfield  {journal}
  {\bibinfo  {journal} {Phys. Rev. Lett.}\ }\textbf {\bibinfo {volume} {93}},\
  \bibinfo {pages} {237204} (\bibinfo {year} {2004})}\BibitemShut {NoStop}%
\bibitem [{\citenamefont {Vidal}\ and\ \citenamefont
  {Dusuel}(2006)}]{J.Vidal_2006}%
  \BibitemOpen
  \bibfield  {author} {\bibinfo {author} {\bibfnamefont {J.}~\bibnamefont
  {Vidal}}\ and\ \bibinfo {author} {\bibfnamefont {S.}~\bibnamefont {Dusuel}},\
  }\bibfield  {title} {\bibinfo {title} {Finite-size scaling exponents in the
  dicke model},\ }\href {https://doi.org/10.1209/epl/i2006-10041-9} {\bibfield
  {journal} {\bibinfo  {journal} {Europhysics Letters}\ }\textbf {\bibinfo
  {volume} {74}},\ \bibinfo {pages} {817} (\bibinfo {year} {2006})}\BibitemShut
  {NoStop}%
\bibitem [{\citenamefont {Torre}\ \emph {et~al.}(2013)\citenamefont {Torre},
  \citenamefont {Diehl}, \citenamefont {Lukin}, \citenamefont {Sachdev},\ and\
  \citenamefont {Strack}}]{DallaTorre2013b}%
  \BibitemOpen
  \bibfield  {author} {\bibinfo {author} {\bibfnamefont {E.~G.~D.}\
  \bibnamefont {Torre}}, \bibinfo {author} {\bibfnamefont {S.}~\bibnamefont
  {Diehl}}, \bibinfo {author} {\bibfnamefont {M.~D.}\ \bibnamefont {Lukin}},
  \bibinfo {author} {\bibfnamefont {S.}~\bibnamefont {Sachdev}},\ and\ \bibinfo
  {author} {\bibfnamefont {P.}~\bibnamefont {Strack}},\ }\bibfield  {title}
  {\bibinfo {title} {Keldysh approach for nonequilibrium phase transitions in
  quantum optics: Beyond the dicke model in optical cavities},\ }\href
  {https://doi.org/10.1103/PhysRevA.87.023831} {\bibfield  {journal} {\bibinfo
  {journal} {Phys. Rev. A}\ }\textbf {\bibinfo {volume} {87}},\ \bibinfo
  {pages} {023831} (\bibinfo {year} {2013})}\BibitemShut {NoStop}%
\bibitem [{\citenamefont {Liberti}\ \emph {et~al.}(2006)\citenamefont
  {Liberti}, \citenamefont {Plastina},\ and\ \citenamefont
  {Piperno}}]{Liberti2006}%
  \BibitemOpen
  \bibfield  {author} {\bibinfo {author} {\bibfnamefont {G.}~\bibnamefont
  {Liberti}}, \bibinfo {author} {\bibfnamefont {F.}~\bibnamefont {Plastina}},\
  and\ \bibinfo {author} {\bibfnamefont {F.}~\bibnamefont {Piperno}},\
  }\bibfield  {title} {\bibinfo {title} {Scaling behavior of the adiabatic
  dicke model},\ }\href {https://doi.org/10.1103/PhysRevA.74.022324} {\bibfield
   {journal} {\bibinfo  {journal} {Phys. Rev. A}\ }\textbf {\bibinfo {volume}
  {74}},\ \bibinfo {pages} {022324} (\bibinfo {year} {2006})}\BibitemShut
  {NoStop}%
\bibitem [{\citenamefont {Liberti}\ \emph {et~al.}(2010)\citenamefont
  {Liberti}, \citenamefont {Piperno},\ and\ \citenamefont
  {Plastina}}]{Liberti2010}%
  \BibitemOpen
  \bibfield  {author} {\bibinfo {author} {\bibfnamefont {G.}~\bibnamefont
  {Liberti}}, \bibinfo {author} {\bibfnamefont {F.}~\bibnamefont {Piperno}},\
  and\ \bibinfo {author} {\bibfnamefont {F.}~\bibnamefont {Plastina}},\
  }\bibfield  {title} {\bibinfo {title} {Finite-size behavior of quantum
  collective spin systems},\ }\href
  {https://doi.org/10.1103/PhysRevA.81.013818} {\bibfield  {journal} {\bibinfo
  {journal} {Phys. Rev. A}\ }\textbf {\bibinfo {volume} {81}},\ \bibinfo
  {pages} {013818} (\bibinfo {year} {2010})}\BibitemShut {NoStop}%
\bibitem [{\citenamefont {Titum}\ and\ \citenamefont
  {Maghrebi}(2020)}]{Titum2020}%
  \BibitemOpen
  \bibfield  {author} {\bibinfo {author} {\bibfnamefont {P.}~\bibnamefont
  {Titum}}\ and\ \bibinfo {author} {\bibfnamefont {M.~F.}\ \bibnamefont
  {Maghrebi}},\ }\bibfield  {title} {\bibinfo {title} {Nonequilibrium
  criticality in quench dynamics of long-range spin models},\ }\href
  {https://doi.org/10.1103/PhysRevLett.125.040602} {\bibfield  {journal}
  {\bibinfo  {journal} {Phys. Rev. Lett.}\ }\textbf {\bibinfo {volume} {125}},\
  \bibinfo {pages} {040602} (\bibinfo {year} {2020})}\BibitemShut {NoStop}%
\bibitem [{\citenamefont {Groszkowski}\ \emph {et~al.}(2022)\citenamefont
  {Groszkowski}, \citenamefont {Koppenh\"ofer}, \citenamefont {Lau},\ and\
  \citenamefont {Clerk}}]{Groszkowski2022}%
  \BibitemOpen
  \bibfield  {author} {\bibinfo {author} {\bibfnamefont {P.}~\bibnamefont
  {Groszkowski}}, \bibinfo {author} {\bibfnamefont {M.}~\bibnamefont
  {Koppenh\"ofer}}, \bibinfo {author} {\bibfnamefont {H.-K.}\ \bibnamefont
  {Lau}},\ and\ \bibinfo {author} {\bibfnamefont {A.~A.}\ \bibnamefont
  {Clerk}},\ }\bibfield  {title} {\bibinfo {title} {Reservoir-engineered spin
  squeezing: Macroscopic even-odd effects and hybrid-systems implementations},\
  }\href {https://doi.org/10.1103/PhysRevX.12.011015} {\bibfield  {journal}
  {\bibinfo  {journal} {Phys. Rev. X}\ }\textbf {\bibinfo {volume} {12}},\
  \bibinfo {pages} {011015} (\bibinfo {year} {2022})}\BibitemShut {NoStop}%
\bibitem [{\citenamefont {Barberena}\ \emph {et~al.}(2019)\citenamefont
  {Barberena}, \citenamefont {Lewis-Swan}, \citenamefont {Thompson},\ and\
  \citenamefont {Rey}}]{Barberena2019}%
  \BibitemOpen
  \bibfield  {author} {\bibinfo {author} {\bibfnamefont {D.}~\bibnamefont
  {Barberena}}, \bibinfo {author} {\bibfnamefont {R.~J.}\ \bibnamefont
  {Lewis-Swan}}, \bibinfo {author} {\bibfnamefont {J.~K.}\ \bibnamefont
  {Thompson}},\ and\ \bibinfo {author} {\bibfnamefont {A.~M.}\ \bibnamefont
  {Rey}},\ }\bibfield  {title} {\bibinfo {title} {Driven-dissipative quantum
  dynamics in ultra-long-lived dipoles in an optical cavity},\ }\href
  {https://doi.org/10.1103/PhysRevA.99.053411} {\bibfield  {journal} {\bibinfo
  {journal} {Phys. Rev. A}\ }\textbf {\bibinfo {volume} {99}},\ \bibinfo
  {pages} {053411} (\bibinfo {year} {2019})}\BibitemShut {NoStop}%
\bibitem [{\citenamefont {Somech}\ and\ \citenamefont
  {Shahmoon}(2022)}]{Somech2022}%
  \BibitemOpen
  \bibfield  {author} {\bibinfo {author} {\bibfnamefont {O.}~\bibnamefont
  {Somech}}\ and\ \bibinfo {author} {\bibfnamefont {E.}~\bibnamefont
  {Shahmoon}},\ }\href {https://doi.org/10.48550/ARXIV.2204.05455} {\bibinfo
  {title} {Quantum entangled states of a classically radiating macroscopic
  spin}} (\bibinfo {year} {2022})\BibitemShut {NoStop}%
\bibitem [{\citenamefont {Dalla~Torre}\ \emph {et~al.}(2013)\citenamefont
  {Dalla~Torre}, \citenamefont {Otterbach}, \citenamefont {Demler},
  \citenamefont {Vuletic},\ and\ \citenamefont {Lukin}}]{DallaTorre2013}%
  \BibitemOpen
  \bibfield  {author} {\bibinfo {author} {\bibfnamefont {E.~G.}\ \bibnamefont
  {Dalla~Torre}}, \bibinfo {author} {\bibfnamefont {J.}~\bibnamefont
  {Otterbach}}, \bibinfo {author} {\bibfnamefont {E.}~\bibnamefont {Demler}},
  \bibinfo {author} {\bibfnamefont {V.}~\bibnamefont {Vuletic}},\ and\ \bibinfo
  {author} {\bibfnamefont {M.~D.}\ \bibnamefont {Lukin}},\ }\bibfield  {title}
  {\bibinfo {title} {Dissipative preparation of spin squeezed atomic ensembles
  in a steady state},\ }\href {https://doi.org/10.1103/PhysRevLett.110.120402}
  {\bibfield  {journal} {\bibinfo  {journal} {Phys. Rev. Lett.}\ }\textbf
  {\bibinfo {volume} {110}},\ \bibinfo {pages} {120402} (\bibinfo {year}
  {2013})}\BibitemShut {NoStop}%
\bibitem [{\citenamefont {Agarwal}\ and\ \citenamefont
  {Puri}(1989)}]{AGARWAL1989}%
  \BibitemOpen
  \bibfield  {author} {\bibinfo {author} {\bibfnamefont {G.}~\bibnamefont
  {Agarwal}}\ and\ \bibinfo {author} {\bibfnamefont {R.}~\bibnamefont {Puri}},\
  }\bibfield  {title} {\bibinfo {title} {Nonequilibrium phase transitions in a
  squeezed cavity and the generation of spin states satisfying uncertainty
  equality},\ }\href
  {https://doi.org/https://doi.org/10.1016/0030-4018(89)90113-2} {\bibfield
  {journal} {\bibinfo  {journal} {Optics Communications}\ }\textbf {\bibinfo
  {volume} {69}},\ \bibinfo {pages} {267} (\bibinfo {year} {1989})}\BibitemShut
  {NoStop}%
\bibitem [{\citenamefont {Guti\'errez-J\'auregui}\ \emph
  {et~al.}(2023)\citenamefont {Guti\'errez-J\'auregui}, \citenamefont
  {Asenjo-Garcia},\ and\ \citenamefont {Agarwal}}]{Gutierrez2023}%
  \BibitemOpen
  \bibfield  {author} {\bibinfo {author} {\bibfnamefont {R.}~\bibnamefont
  {Guti\'errez-J\'auregui}}, \bibinfo {author} {\bibfnamefont {A.}~\bibnamefont
  {Asenjo-Garcia}},\ and\ \bibinfo {author} {\bibfnamefont {G.~S.}\
  \bibnamefont {Agarwal}},\ }\bibfield  {title} {\bibinfo {title} {Dissipative
  stabilization of dark quantum dimers via squeezed vacuum},\ }\href
  {https://doi.org/10.1103/PhysRevResearch.5.013127} {\bibfield  {journal}
  {\bibinfo  {journal} {Phys. Rev. Res.}\ }\textbf {\bibinfo {volume} {5}},\
  \bibinfo {pages} {013127} (\bibinfo {year} {2023})}\BibitemShut {NoStop}%
\bibitem [{\citenamefont {Drummond}(1980)}]{Drummondo1980}%
  \BibitemOpen
  \bibfield  {author} {\bibinfo {author} {\bibfnamefont {P.~D.}\ \bibnamefont
  {Drummond}},\ }\bibfield  {title} {\bibinfo {title} {Observables and moments
  of cooperative resonance fluorescence},\ }\href
  {https://doi.org/10.1103/PhysRevA.22.1179} {\bibfield  {journal} {\bibinfo
  {journal} {Phys. Rev. A}\ }\textbf {\bibinfo {volume} {22}},\ \bibinfo
  {pages} {1179} (\bibinfo {year} {1980})}\BibitemShut {NoStop}%
\bibitem [{\citenamefont {Walls}(1980)}]{Walls1980}%
  \BibitemOpen
  \bibfield  {author} {\bibinfo {author} {\bibfnamefont {D.~F.}\ \bibnamefont
  {Walls}},\ }\bibfield  {title} {\bibinfo {title} {Cooperative fluorescence
  from n coherently driven two-level atoms},\ }\href
  {https://doi.org/10.1088/0022-3700/13/10/008} {\bibfield  {journal} {\bibinfo
   {journal} {Journal of Physics B: Atomic and Molecular Physics}\ }\textbf
  {\bibinfo {volume} {13}},\ \bibinfo {pages} {2001} (\bibinfo {year}
  {1980})}\BibitemShut {NoStop}%
\bibitem [{\citenamefont {Buča}\ and\ \citenamefont
  {Prosen}(2012)}]{Buca_2012}%
  \BibitemOpen
  \bibfield  {author} {\bibinfo {author} {\bibfnamefont {B.}~\bibnamefont
  {Buča}}\ and\ \bibinfo {author} {\bibfnamefont {T.}~\bibnamefont {Prosen}},\
  }\bibfield  {title} {\bibinfo {title} {A note on symmetry reductions of the
  lindblad equation: transport in constrained open spin chains},\ }\href
  {https://doi.org/10.1088/1367-2630/14/7/073007} {\bibfield  {journal}
  {\bibinfo  {journal} {New Journal of Physics}\ }\textbf {\bibinfo {volume}
  {14}},\ \bibinfo {pages} {073007} (\bibinfo {year} {2012})}\BibitemShut
  {NoStop}%
\bibitem [{\citenamefont {Lieu}\ \emph {et~al.}(2020)\citenamefont {Lieu},
  \citenamefont {Belyansky}, \citenamefont {Young}, \citenamefont {Lundgren},
  \citenamefont {Albert},\ and\ \citenamefont {Gorshkov}}]{lieu2020symmetry}%
  \BibitemOpen
  \bibfield  {author} {\bibinfo {author} {\bibfnamefont {S.}~\bibnamefont
  {Lieu}}, \bibinfo {author} {\bibfnamefont {R.}~\bibnamefont {Belyansky}},
  \bibinfo {author} {\bibfnamefont {J.~T.}\ \bibnamefont {Young}}, \bibinfo
  {author} {\bibfnamefont {R.}~\bibnamefont {Lundgren}}, \bibinfo {author}
  {\bibfnamefont {V.~V.}\ \bibnamefont {Albert}},\ and\ \bibinfo {author}
  {\bibfnamefont {A.~V.}\ \bibnamefont {Gorshkov}},\ }\bibfield  {title}
  {\bibinfo {title} {Symmetry breaking and error correction in open quantum
  systems},\ }\href@noop {} {\bibfield  {journal} {\bibinfo  {journal} {Phys.
  Rev. Lett.}\ }\textbf {\bibinfo {volume} {125}},\ \bibinfo {pages} {240405}
  (\bibinfo {year} {2020})}\BibitemShut {NoStop}%
\bibitem [{\citenamefont {Radcliffe}(1971)}]{JMRadcliffe_1971}%
  \BibitemOpen
  \bibfield  {author} {\bibinfo {author} {\bibfnamefont {J.~M.}\ \bibnamefont
  {Radcliffe}},\ }\bibfield  {title} {\bibinfo {title} {Some properties of
  coherent spin states},\ }\href {https://doi.org/10.1088/0305-4470/4/3/009}
  {\bibfield  {journal} {\bibinfo  {journal} {Journal of Physics A: General
  Physics}\ }\textbf {\bibinfo {volume} {4}},\ \bibinfo {pages} {313} (\bibinfo
  {year} {1971})}\BibitemShut {NoStop}%
\bibitem [{\citenamefont {Koppenh\"ofer}\ \emph {et~al.}(2023)\citenamefont
  {Koppenh\"ofer}, \citenamefont {Groszkowski},\ and\ \citenamefont
  {Clerk}}]{Koppenhofer2023}%
  \BibitemOpen
  \bibfield  {author} {\bibinfo {author} {\bibfnamefont {M.}~\bibnamefont
  {Koppenh\"ofer}}, \bibinfo {author} {\bibfnamefont {P.}~\bibnamefont
  {Groszkowski}},\ and\ \bibinfo {author} {\bibfnamefont {A.~A.}\ \bibnamefont
  {Clerk}},\ }\bibfield  {title} {\bibinfo {title} {Squeezed superradiance
  enables robust entanglement-enhanced metrology even with highly imperfect
  readout},\ }\href {https://doi.org/10.1103/PhysRevLett.131.060802} {\bibfield
   {journal} {\bibinfo  {journal} {Phys. Rev. Lett.}\ }\textbf {\bibinfo
  {volume} {131}},\ \bibinfo {pages} {060802} (\bibinfo {year}
  {2023})}\BibitemShut {NoStop}%
\bibitem [{\citenamefont {Muñoz}\ \emph {et~al.}(2019)\citenamefont {Muñoz},
  \citenamefont {Buča}, \citenamefont {Tindall}, \citenamefont
  {González-Tudela}, \citenamefont {Jaksch},\ and\ \citenamefont
  {Porras}}]{munoz2019nonstationary}%
  \BibitemOpen
  \bibfield  {author} {\bibinfo {author} {\bibfnamefont {C.~S.}\ \bibnamefont
  {Muñoz}}, \bibinfo {author} {\bibfnamefont {B.}~\bibnamefont {Buča}},
  \bibinfo {author} {\bibfnamefont {J.}~\bibnamefont {Tindall}}, \bibinfo
  {author} {\bibfnamefont {A.}~\bibnamefont {González-Tudela}}, \bibinfo
  {author} {\bibfnamefont {D.}~\bibnamefont {Jaksch}},\ and\ \bibinfo {author}
  {\bibfnamefont {D.}~\bibnamefont {Porras}},\ }\href@noop {} {\bibinfo {title}
  {Non-stationary dynamics and dissipative freezing in squeezed superradiance}}
  (\bibinfo {year} {2019}),\ \Eprint {https://arxiv.org/abs/1903.05080}
  {arXiv:1903.05080 [quant-ph]} \BibitemShut {NoStop}%
\bibitem [{\citenamefont {Agarwal}\ and\ \citenamefont
  {Puri}(1990)}]{Agarwal1990}%
  \BibitemOpen
  \bibfield  {author} {\bibinfo {author} {\bibfnamefont {G.~S.}\ \bibnamefont
  {Agarwal}}\ and\ \bibinfo {author} {\bibfnamefont {R.~R.}\ \bibnamefont
  {Puri}},\ }\bibfield  {title} {\bibinfo {title} {Cooperative behavior of
  atoms irradiated by broadband squeezed light},\ }\href
  {https://doi.org/10.1103/PhysRevA.41.3782} {\bibfield  {journal} {\bibinfo
  {journal} {Phys. Rev. A}\ }\textbf {\bibinfo {volume} {41}},\ \bibinfo
  {pages} {3782} (\bibinfo {year} {1990})}\BibitemShut {NoStop}%
\bibitem [{\citenamefont {Kuzmich}\ \emph {et~al.}(1997)\citenamefont
  {Kuzmich}, \citenamefont {M\o{}lmer},\ and\ \citenamefont
  {Polzik}}]{Kuzmich1997}%
  \BibitemOpen
  \bibfield  {author} {\bibinfo {author} {\bibfnamefont {A.}~\bibnamefont
  {Kuzmich}}, \bibinfo {author} {\bibfnamefont {K.}~\bibnamefont {M\o{}lmer}},\
  and\ \bibinfo {author} {\bibfnamefont {E.~S.}\ \bibnamefont {Polzik}},\
  }\bibfield  {title} {\bibinfo {title} {Spin squeezing in an ensemble of atoms
  illuminated with squeezed light},\ }\href
  {https://doi.org/10.1103/PhysRevLett.79.4782} {\bibfield  {journal} {\bibinfo
   {journal} {Phys. Rev. Lett.}\ }\textbf {\bibinfo {volume} {79}},\ \bibinfo
  {pages} {4782} (\bibinfo {year} {1997})}\BibitemShut {NoStop}%
\bibitem [{\citenamefont {Susskind}\ and\ \citenamefont
  {Glogower}(1964)}]{Susskind1964}%
  \BibitemOpen
  \bibfield  {author} {\bibinfo {author} {\bibfnamefont {L.}~\bibnamefont
  {Susskind}}\ and\ \bibinfo {author} {\bibfnamefont {J.}~\bibnamefont
  {Glogower}},\ }\bibfield  {title} {\bibinfo {title} {Quantum mechanical phase
  and time operator},\ }\href
  {https://doi.org/10.1103/PhysicsPhysiqueFizika.1.49} {\bibfield  {journal}
  {\bibinfo  {journal} {Physics Physique Fizika}\ }\textbf {\bibinfo {volume}
  {1}},\ \bibinfo {pages} {49} (\bibinfo {year} {1964})}\BibitemShut {NoStop}%
\bibitem [{\citenamefont {Wineland}\ \emph {et~al.}(1992)\citenamefont
  {Wineland}, \citenamefont {Bollinger}, \citenamefont {Itano}, \citenamefont
  {Moore},\ and\ \citenamefont {Heinzen}}]{Wineland1992}%
  \BibitemOpen
  \bibfield  {author} {\bibinfo {author} {\bibfnamefont {D.~J.}\ \bibnamefont
  {Wineland}}, \bibinfo {author} {\bibfnamefont {J.~J.}\ \bibnamefont
  {Bollinger}}, \bibinfo {author} {\bibfnamefont {W.~M.}\ \bibnamefont
  {Itano}}, \bibinfo {author} {\bibfnamefont {F.~L.}\ \bibnamefont {Moore}},\
  and\ \bibinfo {author} {\bibfnamefont {D.~J.}\ \bibnamefont {Heinzen}},\
  }\bibfield  {title} {\bibinfo {title} {Spin squeezing and reduced quantum
  noise in spectroscopy},\ }\href {https://doi.org/10.1103/PhysRevA.46.R6797}
  {\bibfield  {journal} {\bibinfo  {journal} {Phys. Rev. A}\ }\textbf {\bibinfo
  {volume} {46}},\ \bibinfo {pages} {R6797} (\bibinfo {year}
  {1992})}\BibitemShut {NoStop}%
\bibitem [{\citenamefont {Puri}\ and\ \citenamefont
  {Lawande}(1979)}]{@Puri1979}%
  \BibitemOpen
  \bibfield  {author} {\bibinfo {author} {\bibfnamefont {R.}~\bibnamefont
  {Puri}}\ and\ \bibinfo {author} {\bibfnamefont {S.}~\bibnamefont {Lawande}},\
  }\bibfield  {title} {\bibinfo {title} {Exact steady-state density operator
  for a collective atomic system in an external field},\ }\href
  {https://doi.org/https://doi.org/10.1016/0375-9601(79)90003-3} {\bibfield
  {journal} {\bibinfo  {journal} {Physics Letters A}\ }\textbf {\bibinfo
  {volume} {72}},\ \bibinfo {pages} {200} (\bibinfo {year} {1979})}\BibitemShut
  {NoStop}%
\bibitem [{\citenamefont {Kilin}(1980)}]{Kilin1980}%
  \BibitemOpen
  \bibfield  {author} {\bibinfo {author} {\bibfnamefont {S.~J.}\ \bibnamefont
  {Kilin}},\ }\bibfield  {title} {\bibinfo {title} {Cooperative resonance
  fluorescence and atomic interactions},\ }\href
  {https://doi.org/10.1088/0022-3700/13/13/023} {\bibfield  {journal} {\bibinfo
   {journal} {Journal of Physics B: Atomic and Molecular Physics}\ }\textbf
  {\bibinfo {volume} {13}},\ \bibinfo {pages} {2653} (\bibinfo {year}
  {1980})}\BibitemShut {NoStop}%
\bibitem [{\citenamefont {Bonifacio}\ \emph {et~al.}(1971)\citenamefont
  {Bonifacio}, \citenamefont {Schwendimann},\ and\ \citenamefont
  {Haake}}]{Bonifacio1971}%
  \BibitemOpen
  \bibfield  {author} {\bibinfo {author} {\bibfnamefont {R.}~\bibnamefont
  {Bonifacio}}, \bibinfo {author} {\bibfnamefont {P.}~\bibnamefont
  {Schwendimann}},\ and\ \bibinfo {author} {\bibfnamefont {F.}~\bibnamefont
  {Haake}},\ }\bibfield  {title} {\bibinfo {title} {Quantum statistical theory
  of superradiance. i},\ }\href {https://doi.org/10.1103/PhysRevA.4.302}
  {\bibfield  {journal} {\bibinfo  {journal} {Phys. Rev. A}\ }\textbf {\bibinfo
  {volume} {4}},\ \bibinfo {pages} {302} (\bibinfo {year} {1971})}\BibitemShut
  {NoStop}%
\bibitem [{\citenamefont {Hannukainen}\ and\ \citenamefont
  {Larson}(2018)}]{Hannukainen2018}%
  \BibitemOpen
  \bibfield  {author} {\bibinfo {author} {\bibfnamefont {J.}~\bibnamefont
  {Hannukainen}}\ and\ \bibinfo {author} {\bibfnamefont {J.}~\bibnamefont
  {Larson}},\ }\bibfield  {title} {\bibinfo {title} {Dissipation-driven quantum
  phase transitions and symmetry breaking},\ }\href
  {https://doi.org/10.1103/PhysRevA.98.042113} {\bibfield  {journal} {\bibinfo
  {journal} {Phys. Rev. A}\ }\textbf {\bibinfo {volume} {98}},\ \bibinfo
  {pages} {042113} (\bibinfo {year} {2018})}\BibitemShut {NoStop}%
\bibitem [{\citenamefont {Link}\ \emph {et~al.}(2019)\citenamefont {Link},
  \citenamefont {Luoma},\ and\ \citenamefont {Strunz}}]{Link2019}%
  \BibitemOpen
  \bibfield  {author} {\bibinfo {author} {\bibfnamefont {V.}~\bibnamefont
  {Link}}, \bibinfo {author} {\bibfnamefont {K.}~\bibnamefont {Luoma}},\ and\
  \bibinfo {author} {\bibfnamefont {W.~T.}\ \bibnamefont {Strunz}},\ }\bibfield
   {title} {\bibinfo {title} {Revealing the nature of nonequilibrium phase
  transitions with quantum trajectories},\ }\href
  {https://doi.org/10.1103/PhysRevA.99.062120} {\bibfield  {journal} {\bibinfo
  {journal} {Phys. Rev. A}\ }\textbf {\bibinfo {volume} {99}},\ \bibinfo
  {pages} {062120} (\bibinfo {year} {2019})}\BibitemShut {NoStop}%
\end{thebibliography}%
\appendix

\section{First eigenvalue of squeezed superradiance steady state}\label{sec:AppFirst}
Here we calculate $\lambda_0$, the first eigenvalue of $\hat{\rho}_{\text{ss}}^{\text{I}}$, which satisfies [see Eq.~(\ref{eqn:OddNSSSpectrum})]
\begin{align}\begin{split}\label{eqn:AppOddNPure}
    \bigg(\delta\hat{n}-\frac{i\zeta N\sin\hat{\phi}}{2}\bigg)\bigg(\delta\hat{n}+\frac{i\zeta N\sin\hat{\phi}}{2}\bigg)\ket{\lambda_0}&=\frac{\ket{\lambda_0}}{\lambda_0}.
\end{split}\end{align}
This can be rewritten as $\hat{O}^\dagger\hat{O}\ket{\lambda_0}=\lambda_0^{-1}\ket{\lambda_0}$, where 
\begin{equation}
    \hat{O}=\delta\hat{n}+\frac{i\zeta N\sin\hat{\phi}}{2}.
\end{equation}
To proceed, we notice that $\hat{O}$ satisfies $\hat{R}^\dagger\hat{O}\hat{R}=\hat{O}^\dagger$, where $\hat{R}=e^{i\pi\delta\hat{n}}$. Multiplying Eq.~(\ref{eqn:AppOddNPure}) by $\hat{R}^\dagger\hat{O}$ on the left and using the relation $\hat{R}^\dagger\hat{O}=\hat{O}^\dagger\hat{R}^\dagger$ repeatedly, we arrive at
\begin{equation}
\hat{O}^\dagger\hat{O}\hat{R}^\dagger\hat{O}\ket{\lambda_0}=\lambda_0^{-1}\hat{R}^\dagger\hat{O}\ket{\lambda_0},
\end{equation}
which indicates that $\hat{R}^\dagger\hat{O}\ket{\lambda_0}$ is also an eigenstate of $\hat{O}^\dagger\hat{O}$ with the same eigenvalue $\lambda_0^{-1}$. Since the approximate dark state is unique, the corresponding eigenstate must be non-degenerate and must satisfy
\begin{equation}
    \hat{O}\ket{\lambda_0}=\beta\hat{R}\ket{\lambda_0},
\end{equation}
where $\beta$ is a proportionality constant. Taking the norm on both sides and using Eq.~(\ref{eqn:AppOddNPure}) we find that $|\beta|^2=\lambda_0^{-1}$. In terms of the wavefunction $\lambda_0(\phi)=\braket{\phi|\lambda_0}$, this equation becomes
\begin{equation}
    i\bigg(\partial_\phi+\frac{\zeta N\sin\phi}{2}\bigg)\lambda_0(\phi)=\beta\lambda_0(\phi+\pi),
\end{equation}
since $\braket{\phi|\hat{R}|\lambda_0}=\braket{\phi+\pi|\lambda_0}$. This can be rewritten in integral form
\begin{equation}\label{eqn:AppFirstEigEquation}
    \lambda_0(\phi)=e^{\frac{\zeta N\cos\phi}{2}}\bigg[\lambda_0(0)-i\beta\int_0^{\phi}\frac{\lambda_0(\phi'+\pi)}{e^{\frac{\zeta N\cos\phi'}{2}}}\,d\phi'\bigg],
\end{equation}
where $\lambda_0(0)$ is an integration constant denoting the value of $\lambda_0(\phi)$ at $\phi=0$. Since $N$ is odd, the spectrum of $\delta\hat{n}=\hat{n}-N/2$ is the half-integers, which implies that $\lambda_(\phi)$ is antiperiodic and therefore $\lambda(2\pi)=-\lambda(0)$. Setting $\phi=2\pi$ in the previous equation leads to
\begin{equation}
    -2\lambda_0(0)=-i\beta\int_0^{2\pi}e^{-\frac{\zeta N\cos\phi}{2}}\lambda_0(\phi+\pi)\,d\phi.
\end{equation}
To lowest order in $\beta$ we can replace $\lambda_0(\phi+\pi)$ in the integral by $\lambda_0(0)\exp[\zeta N\cos(\phi+\pi)/2]$. The factors of $\lambda_0(0)$ then cancel, which gives us an equation for $\beta$
\begin{equation}
    2=i\beta\int_0^{2\pi}e^{-\zeta N\cos\phi}\,d\phi=2\pi i \beta I_0(\zeta N),
\end{equation}
where $I_0(x)$ is the $0^{\text{th}}$ order modified Bessel function of the first kind. Hence, we arrive at
\begin{equation}
    \lambda_0=|\beta|^{-2}=\pi^2 \big[I_0(\zeta N)\big]^2,
\end{equation}
which is Eq.~(\ref{eqn:OddNEigenvalue}) in the main text. In Fig.~\ref{fig:App1}(a) we compare this analytical result against an exact numerical calculation using Eq.~(\ref{eqn:Apprhosdm}) for $N=1001$. We find good agreement when $\lambda_0\gtrsim 1$.

\section{Bulk integral of squeezed superradiance}\label{sec:AppBulk}
\begin{figure*}
    \centering
    \includegraphics[width=0.98\textwidth]{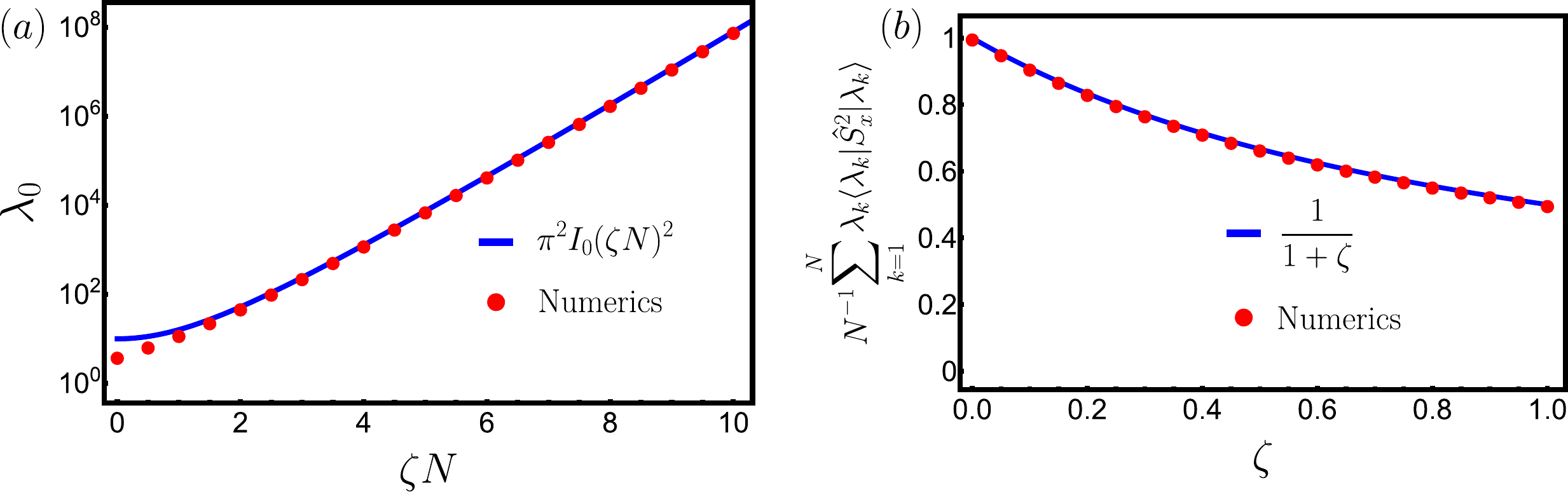}
    \caption{(a) First eigenvalue of $\hat{\rho}_{\text{ss}}^{\text{I}}$ as a function of $\zeta N$. We show the analytical approximation $\pi^2I_0(\zeta N)^2$ (blue line) and the exact numerical result for $N=1001$ (red dots). (b) Bulk sum of $\hat{S}_x^2$ as a function of $\zeta$. We show the analytical approximation $(1+\zeta)^{-1}$ (blue line) and the exact numerical result for $N=1001$ (red dots). }
    \label{fig:App1}
\end{figure*}

Here we calculate the bulk sum that appears in Eq.~(\ref{eqn:SDMBulkSum}),
\begin{equation}
\mathrm{Tr}(\hat{\rho}_{\text{ss}}^{\text{I}}\hat{S}_x^2)=\lambda_0\braket{\lambda_0|\hat{S}_x^2|\lambda_0}+\boxed{\sum_{k=1}^N\lambda_k\braket{\lambda_k|\hat{S}_x^2|\lambda_k}},
\end{equation}
where $\hat{\rho}_{\text{ss}}^{\text{I}}$ is the (unnormalized) steady state of squeezed superradiance (SSR), given by
\begin{equation}\label{eqn:Apprhosdm}
    \hat{\rho}_{\text{ss}}^{\text{I}}=\big[(\hat{S}_x+i\zeta\hat{S}_y)(\hat{S}_x-i\zeta\hat{S}_y)\big]^{-1},
\end{equation}
and $\ket{\lambda_k}$ is its $k^{\text{th}}$ eigenstate in order of decreasing magnitude. The boxed sum can be calculated semiclassically from the exact expression for the steady state. To do so, we perform the semiclassical replacements $\hat{S}_z\to N\cos\theta/2$ and $\hat{S}^-\to N \sin\theta e^{i\phi}/2$, where $\theta$ and $\phi$ are spherical coordinates angles, and approximate the sum by an integral over phase space
\begin{equation}
    \sum_{k=1}^N\lambda_k\braket{\lambda_k|\hat{S}_x^2|\lambda_k}\approx \frac{N}{4\pi}\int  \frac{\cos\phi^2\sin\theta\,d\theta d\phi}{(\cos\phi^2+\zeta^2\sin\phi^2)}.
\end{equation}
The numerator in the previous equation comes from the average of $\hat{S}_x^2$ over eigenstates in the sum, while the denominator comes from $\hat{\rho}_{\text{ss}}^{\text{I}}$, which corresponds to the $\lambda_k$ factors in the sum. The integral can be done exactly, and leads to
\begin{equation}
    \sum_{k=1}^N\lambda_k\braket{\lambda_k|\hat{S}_x^2|\lambda_k}\approx \frac{N}{1+\zeta}.
\end{equation}
We compare this analytical result against a numerical calculation using Eq.~(\ref{eqn:Apprhosdm}) in Fig.~\ref{fig:App1}(b), where we find excellent agreement over the whole parameter region $\zeta\in [0,1]$.

\section{First eigenvalue of driven superradiance}\label{sec:AppFirstCRF}
Here we calculate $\tilde{\mu}_0$, proportional to the first eigenvalue of $\hat{\rho}_{\text{ss}}^{\text{II}}$, which satisfies [see Eq.~(\ref{eqn:CRFBosonEquation})]
\begin{equation}\label{eqn:AppCRFBosonEquation}
        \big(\hat{y}-i\hat{q}^2-i\eta\big)\big(\hat{y}+i\hat{q}^2+i\eta\big)\ket{\mu_0}=\frac{1}{\tilde{\mu}_0}\ket{\mu_0},
\end{equation}
This can be rewritten as $\hat{M}^\dagger\hat{M}\ket{\mu_0}=\tilde{\mu}_0^{-1}\ket{\mu_0}$, where 
\begin{equation}
    \hat{M}=\hat{y}+i\hat{q}^2+i\eta.
\end{equation}
To proceed we notice that $\hat{M}$ satisfies $\hat{S}^\dagger \hat{M}\hat{S}=-\hat{M}^\dagger$, where $\hat{S}=\exp[i\pi(\hat{x}^2+\hat{p}^2)/2]$. Multiplying Eq.~(\ref{eqn:AppCRFBosonEquation}) by $\hat{S}^\dagger\hat{M}$ on the left and using the relation $\hat{S}^\dagger \hat{M}=-\hat{M}^\dagger \hat{S}^\dagger$ repeatedly, we arrive at
\begin{equation}
    \hat{M}^\dagger\hat{M}\hat{S}^\dagger\hat{M}\ket{\mu_0}=\tilde{\mu}_0^{-1}\hat{S}^\dagger\hat{M}\ket{\mu_0},
\end{equation}
which indicates that $\hat{S}^\dagger\hat{M}\ket{\mu_0}$ is also an eigenstate of $\hat{M}^\dagger\hat{M}$ with the same eigenvalue $\tilde{\mu}_0^{-1}$. Since the approximate dark state is unique, the corresponding eigenstate must be non-degenerate and it must satisfy
\begin{equation}
    \hat{M}\ket{\mu_0}=\gamma \hat{S}\ket{\mu_0},
\end{equation}
where $\gamma$ is a proportionality constant. Taking the norm on both sides and using Eq.~(\ref{eqn:AppCRFBosonEquation}) we find that $|\gamma|^2=\tilde{\mu}_0^{-1}$. In terms of the wavefunction $\mu_0(q)=\braket{q|\mu_0}$, this equation becomes
\begin{equation}
    i(\partial_q+q^2+\eta)\mu_0(q)=\gamma \mu_0(-q).
\end{equation}
since $\braket{q|\hat{S}|\mu_0}=\braket{-q|\mu_0}$. This can be rewritten in integral form
\begin{equation}
    \mu_0(q)=e^{-\frac{q^3}{3}-\eta q}\bigg(\mu_0(0)-i\gamma\int_0^q \frac{\mu_0(-q')\,dq'}{e^{-\frac{q'^3}{3}-\eta q'}}\bigg),
\end{equation}
where $\mu_0(0)$ is an integration constant denoting the value of $\mu_0(q)$ at $q=0$. Normalizability as $q\to -\infty$ requires that the term in parentheses approach 0 in this limit. This gives us the following condition
\begin{equation}
    \mu_0(0)=i\gamma\int_0^{-\infty}e^{\frac{q^3}{3}+\eta q}\mu_0(-q)\,dq.
\end{equation}
To lowest order in $\gamma$ we can replace $\mu_0(-q)$ in the integral by $\mu_0(0)\exp[-(-q)^3/3-\eta (-q)]$. The factors of $\mu_0(0)$ then cancel, leaving an equation for $\gamma$,
\begin{equation}
    1=i\gamma \int_0^{-\infty}e^{\frac{2q^3}{3}+2\eta q}\,dq.
\end{equation}
Hence, we arrive at
\begin{equation}
    \tilde{\mu}_0=|\gamma|^{-2}=\bigg[\int_0^{\infty}e^{-\frac{2q^3}{3}-2\eta q}\,dq\bigg]^2,
\end{equation}
which is Eq.~(\ref{eqn:CRFLowestEigenvalue}) in the main text. In Fig.~\ref{fig:App2}(a) we compare this analytical result against an exact numerical calculation using Eq.~(\ref{eqn:Apprhocrf}) for $N=10001$. We find good agreement when $\eta\lesssim -0.5$.

\section{Bulk integral of driven superradiance}\label{sec:AppBulkCRf}
\begin{figure*}
    \centering
    \includegraphics[width=0.98\textwidth]{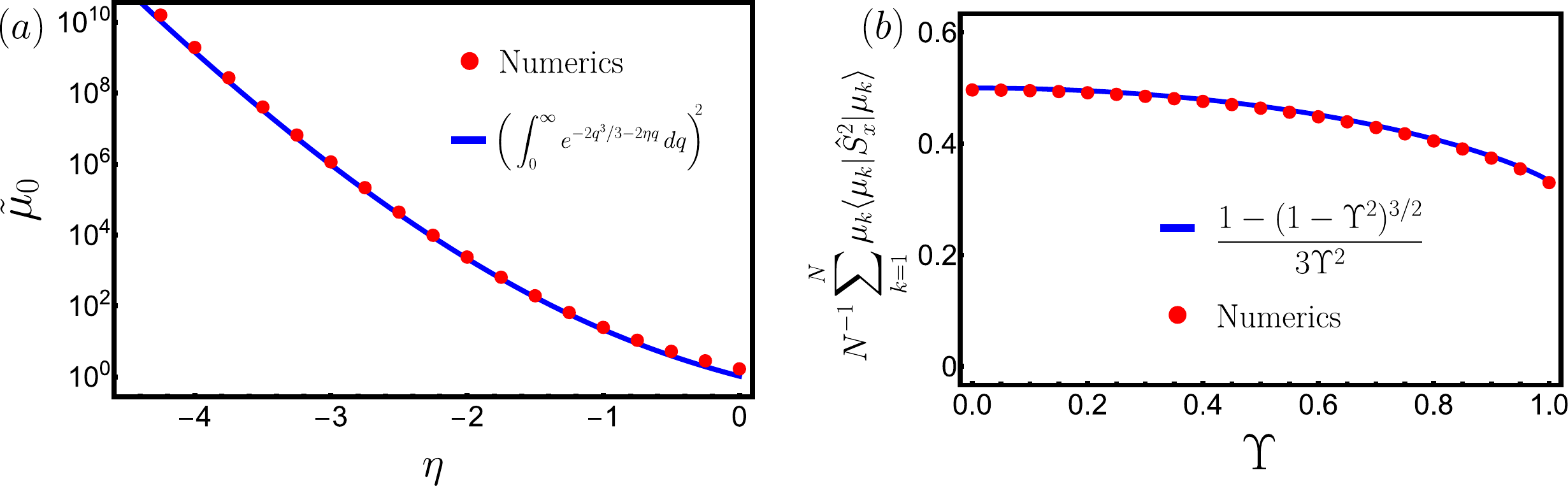}
    \caption{(a) First eigenvalue of $\hat{\rho}_{\text{ss}}^{\text{II}}$ as a function of $\eta$. We show the analytical approximation (blue line) and the exact numerical result for $N=10001$ (red dots). (b) Bulk sum of $\hat{S}_x^2$ as a function of $\Upsilon$. We show the analytical approximation (blue line) and the exact numerical result for $N=1001$ (red dots). }
    \label{fig:App2}
\end{figure*}

Here we calculate the sum that appears in Eq.~(\ref{eqn:CRFBulkSum}),
\begin{equation}
\mathrm{Tr}(\hat{\rho}_{\text{ss}}^{\text{II}}\hat{S}_x^2)=\mu_0\braket{\mu_0|\hat{S}_x^2|\mu_0}+\boxed{\sum_{k=1}^N\mu_k\braket{\mu_k|\hat{S}_x^2|\mu_k}},
\end{equation}
where $\hat{\rho}_{\text{ss}}^{\text{II}}$ is the (unnormalized) steady state of the driven superradiance, given by
\begin{equation}\label{eqn:Apprhocrf}
    \hat{\rho}_{\text{ss}}^{\text{II}}=\Bigg[\bigg(\hat{S}^++\frac{iN\Upsilon}{2}\bigg)\bigg(\hat{S}^-+\frac{iN\Upsilon}{2}\bigg)\Bigg]^{-1},
\end{equation}
and $\ket{\mu_k}$ is its $k^{\text{th}}$ eigenstate in order of decreasing magnitude. The boxed sum can be calculated semiclassically from the exact expression for the steady state, which leads to
\begin{equation}
    \sum_{k=1}^N\mu_k\braket{\mu_k|\hat{S}_x^2|\mu_k}\approx \frac{N}{4\pi}\int  \frac{\sin\theta^2\cos\phi^2\sin\theta\,d\theta d\phi}{\big|\sin\theta\, e^{i\phi}+i\Upsilon\big|^2}.
\end{equation}
The origin of the terms is similar to that in the SDM, where the numerator corresponds to the average of $\hat{S}_x^2$ and the denominator to the $\mu_k$ factors. Evaluation of the integral (by e.g. contour integration) leads to
\begin{equation}
    \sum_{k=1}^N\mu_k\braket{\mu_k|\hat{S}_x^2|\mu_k}\approx \frac{N}{3}\bigg[\frac{1-(1-\Upsilon^2)^{3/2}}{\Upsilon^2}\bigg].
\end{equation}
We compare this analytical result against a numerical calculation using Eq.~(\ref{eqn:Apprhocrf}) in Fig.~\ref{fig:App2}(b), where we find excellent agreement over the whole parameter region $\Upsilon\in [0,1]$. Close to the transition point we set $\Upsilon=1$, which gives the value $N/3$ for the sum and which leads to Eq.~(\ref{eqn:CRFSxVariance}). 

\section{Contrast in driven superradiance}\label{sec:AppContrast}
We calculate the values of $\braket{\hat{S}_y}$, $\braket{\hat{S}_z}$ and $\braket{\hat{S}_x^2}$ in the steady state $\hat{\rho}_{\text{ss}}^{\text{II}}$ given by Eq.~(\ref{eqn:Apprhocrf}) up to the transition point $\Upsilon\approx 1$. We begin with $\braket{\hat{S}_z}$, then continue with $\braket{\hat{S}_y}$, which is trickier and finalize with $\braket{\hat{S}_x^2}$, which is the trickiest.
\subsection{Z component}
By definition,
\begin{equation}
    \braket{\hat{S}_z}=\frac{\mathrm{Tr}(\hat{\rho}_{\text{ss}}^{\text{II}}\hat{S}_z)}{\mathrm{Tr}(\hat{\rho}_{\text{ss}}^{\text{II}})}.
\end{equation}
We assume the bosonic approximation for $\hat{\rho}_{\text{ss}}^{\text{II}}$ given in Eq.~(\ref{eqn:CRFBosonEquation}) and write $\hat{S}_z\approx N^{2/3}\hat{q}/2^{1/3}$ using Eq.~(\ref{eqn:CRFLinearization}) (with $\alpha=\pi/2$) and Eq.~(\ref{eqn:Rescaling}) to get
\begin{align}\begin{split}\label{eqn:AppSz}
    \braket{\hat{S}_z}&=\bigg(\frac{N^{2/3}}{2^{1/3}}\bigg)\frac{\overbrace{\mathrm{Tr}\bigg[\bigg(\frac{1}{\hat{q}^2+\eta-i\hat{y}}\bigg)\bigg(\frac{1}{\hat{q}^2+\eta+i\hat{y}}\bigg)\hat{q}\bigg]}^{\text{Num}}}{\underbrace{\mathrm{Tr}\bigg[\bigg(\frac{1}{\hat{q}^2+\eta-i\hat{y}}\bigg)\bigg(\frac{1}{\hat{q}^2+\eta+i\hat{y}}\bigg)\bigg]}_{\text{Den}}}
\end{split}\end{align}
The numerator can be rewritten as
\begin{equation}
    \text{Num}=\int_0^{\infty}\,ds\,dr \mathrm{Tr}\Big[e^{-s(\hat{q}^2+\eta-i\hat{y})}e^{-r(\hat{q}^2+\eta+i\hat{y})}\hat{q}\Big],
\end{equation}
where both $s$ and $r$ are integrated over the range $(0,\infty)$. Since each exponential involves $\hat{y}$ and $\hat{q}$ at most to quadratic order, we can express this as
\begin{equationS}
    \text{Num}=\int_0^{\infty}\,ds\,dr \mathrm{Tr}\Big[&\Big(e^{is\hat{y}-\eta s}\Big)\Big(e^{-\hat{q}^2 s-\hat{q}s^2-s^3/3}\Big)\\
    &\Big(e^{-\hat{q}^2 r-\hat{q}r^2-r^3/3}\Big)\Big(e^{-ir\hat{y}-\eta r}\Big)\hat{q}\Big]\\[5pt]
    =\int_0^{\infty}\,ds\,dr \mathrm{Tr}\Big[&e^{-\hat{q}^2(r+s)-\hat{q}(r^2+s^2)-(s^3+r^3)/3}\\
    &e^{-\eta(r+s)}(\hat{q}+r)e^{i(s-r)\hat{y}}\Big],
\end{equationS}
where $\hat{q}$ has been shifted to $\hat{q}+r$ because $e^{-i\hat{y}r}\hat{q}=(\hat{q}+r)e^{-i\hat{y}r}$. Introducing two resolutions of the identity, one with respect to $\hat{y}$ eigenstates and the other with respect to $\hat{q}$ eigenstates, we can express the numerator as
\begin{equationS}\label{eqn:AppSzNum}
    \text{Num}=\int_{-\infty}^{\infty}&\frac{dq\,dy}{2\pi}\int_0^{\infty}\,ds\,dr \,e^{-q^2(r+s)-q(r^2+s^2)}\\
    &e^{-(s^3+r^3)/3-\eta (r+s)}(q+r)e^{i(s-r)y}\\[5pt]
    =\int_{-\infty}^{\infty}&dq\int_0^{\infty}\,ds\,e^{-2q^2s-2qs-2s^3/2-2\eta s}(q+s).
\end{equationS}
Integrating with respect to $q$ and changing variables from $s$ to $v=\sqrt{s}$ we obtain
\begin{equation}
    \text{Num}=\sqrt{2\pi}\int_0^{\infty}\,dv\,\Big( e^{-v^6/6-2\eta v^2}\Big) \frac{v^2}{2}.
\end{equation}
Similar manipulations for the denominator in Eq.~(\ref{eqn:AppSz}) give the same integral as in Eq.~(\ref{eqn:AppSzNum}) but omitting the $(q+s)$ factor. This leads to
\begin{equation}
    \text{Den}=\sqrt{2\pi}\int_0^{\infty}\,dv\,\Big( e^{-v^6/6-2\eta v^2}\Big),
\end{equation}
and finally to
\begin{equation}\label{eqn:AppSzFin}
    \braket{\hat{S}_z}=\bigg(\frac{N}{4}\bigg)^{2/3}\frac{\int_0^\infty\,dv\, e^{-v^6/6-2\eta v^2}v^2}{\int_0^\infty\,dv\, e^{-v^6/6-2\eta v^2}},
\end{equation}
which is the first line of Eq.~(\ref{eqn:CRFContrast}) in the main text. We compare this against exact numerical simulations in Fig.~\ref{fig:AppUniformApproximations}(a), showing very good agreement over an extended range of $\delta\Upsilon$, especially at larger values of $N$.
\subsection{Y Component}
By definition,

\begin{equation}\label{eqn:AppSy}
    \braket{\delta\hat{S}_y}=\frac{\mathrm{Tr}[\hat{\rho}_{\text{ss}}^{\text{II}}\delta\hat{S}_y]}{\mathrm{Tr}(\hat{\rho}_{\text{ss}}^{\text{II}})},
\end{equation}
where $\delta\hat{S}_y=\hat{S}_y-N/2$, and we subtract $N/2$ since we know that the Bloch vector is polarized along the $+y$ direction. Naive application of the boson approximation will lead to a divergent result however, so care is needed when computing this observable. To get a convergent result, we re-express the numerator of Eq.~(\ref{eqn:AppSy}) as
\begin{equationS}
    \mathrm{Tr}[\hat{\rho}_{\text{ss}}^{\text{II}}(\eta)\delta\hat{S}_y]&=\overbrace{\mathrm{Tr}\{[\hat{\rho}_{\text{ss}}^{\text{II}}(\eta)-\hat{\rho}_{\text{ss}}^{\text{II}}(0)]\delta\hat{S}_y\}}^{A}\\[5pt]
    &+\underbrace{\mathrm{Tr}[\hat{\rho}_{\text{ss}}^{\text{II}}(0)\delta\hat{S}_y]}_{B},
\end{equationS}
where we have decided to make explicit the dependence of $\hat{\rho}_{\text{ss}}^{\text{II}}(\eta)$ on $\eta$ and $\hat{\rho}_{\text{ss}}^{\text{II}}(0)$ is the steady state at the critical point. We can calculate $B$ using semiclassical analysis
\begin{equationS}
    B&=\frac{N}{4\pi}\int \sin\theta\,d\theta\,d\phi\frac{\frac{N}{2}(\sin\theta\sin\phi-1)}{\big(\frac{N}{2}\big)^2|\sin\theta e^{i\phi}-1|^2}\\
    &=-2.
\end{equationS}
We can calculate $A$ using the boson approximation
  \begin{equationS}
        A&=-\frac{1}{(N/4)^{2/3}}\mathrm{Tr}\Bigg\{\Bigg[\bigg(\frac{1}{\hat{q}^2+\eta-i\hat{y}}\bigg)\bigg(\frac{1}{\hat{q}^2+\eta+i\hat{y}}\bigg)\\[5pt]
    &-\bigg(\frac{1}{\hat{q}^2-i\hat{y}}\bigg)\bigg(\frac{1}{\hat{q}^2+i\hat{y}}\bigg)\Bigg]\frac{(2N)^{1/3}\hat{q}^2}{2}\Bigg\}
    \end{equationS}

\begin{figure*}
    \centering
    \includegraphics[width=0.98\textwidth]{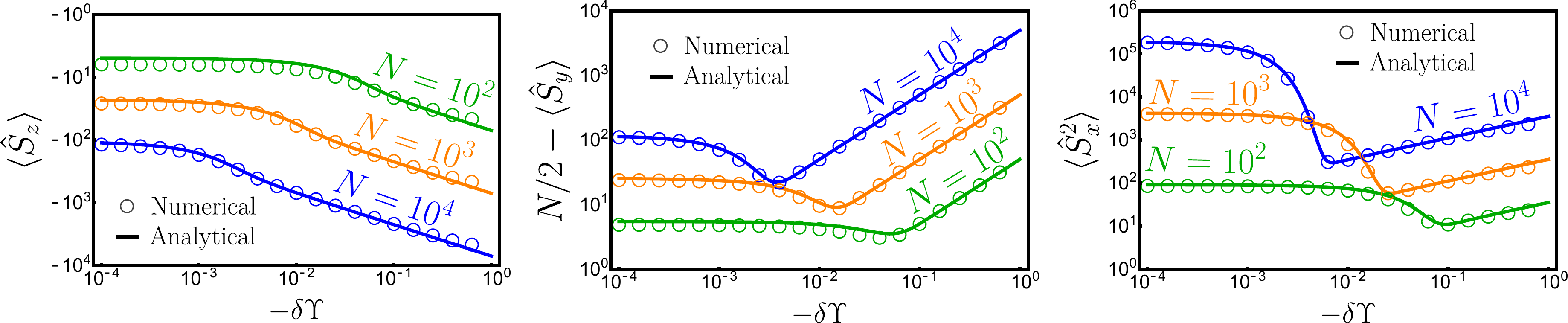}
    \caption{Comparison between numerical solution of the master equation of driven superradiance (empty circles) and analytical formulas (solid lines) for $N=10^2$ (green), $N=10^3$ (orange) and $N=10^4$ (blue) as a function of the distance (from below) to the critical point $-\delta\Upsilon=1-\Upsilon$. (a) Plot of $\braket{\hat{S}_z}$ and Eq.~(\ref{eqn:AppSzFin}). (b) Plot of $N/2-\braket{\hat{S}_y}$ and Eq.~(\ref{eqn:AppSyFin}). (c) Plot of $\braket{\hat{S}_x^2}$ and Eq.~(\ref{eqn:AppSx2Fin}). There is more discrepancy towards $-\delta\Upsilon \sim O(1)$, where the mean field approximation gives correct results anyway. As should be expected, agreement gets better with larger $N$ for all quantities.}
    \label{fig:AppUniformApproximations}
\end{figure*}

We have used $\delta\hat{S}_y=-(\hat{x}^2+\hat{p}^2-1)/2\approx -\hat{p}^2/2=-(2N)^{1/3}\hat{q}^2/2$ to get the leading contribution. Application of the same techniques used to get $\braket{\hat{S}_z}$ leads to
\begin{equation}
    A=\frac{(2N)^{1/3}/2}{(N/4)^{2/3}}\bigg[\eta\sqrt{2\pi}\int_0^{\infty} e^{-v^6/6-2\eta v^2}\,dv\bigg].
\end{equation}
Similarly
\begin{equationS}
    \mathrm{Tr}[\hat{\rho}_{\text{ss}}^{\text{II}}(\eta)]&=\frac{1}{(N/4)^{2/3}}\mathrm{Tr}\bigg[\bigg(\frac{1}{\hat{q}^2+\eta-i\hat{y}}\bigg)\bigg(\frac{1}{\hat{q}^2+\eta+i\hat{y}}\bigg)\bigg]\\[5pt]
    &=\frac{1}{(N/4)^{2/3}}\bigg[\sqrt{2\pi}\int_0^{\infty} e^{-v^6/6-2\eta v^2}\,dv\bigg].
\end{equationS}
Putting these results together
\begin{equation}\label{eqn:AppSyFin}
    \braket{\hat{S}_y}-\frac{N}{2}=\frac{(2N)^{1/3}\eta}{2}-\sqrt{\frac{2}{\pi}}\frac{(N/4)^{2/3}}{\int_0^\infty e^{-v^6/6-2\eta v^2}\,dv},
\end{equation}
which is the second line of Eq.~(\ref{eqn:CRFContrast}). We compare this against exact numerical simulations in Fig.~\ref{fig:AppUniformApproximations}(b), showing very good agreement over an extended range of $\delta\Upsilon$, especially at larger $N$.
\subsection{Variance along X}\label{sec:AppContrastVariance}
Here we provide another approximation for $\braket{\hat{S}_x^2}$ which is valid over a larger range of $\delta\Upsilon$, not only near its minimum value. By definition
\begin{equation}\label{eqn:AppSx2}
    \braket{\hat{S}_x^2}=\frac{\mathrm{Tr}[\hat{\rho}_{\text{ss}}^{\text{II}}\hat{S}_x^2]}{\mathrm{Tr}(\hat{\rho}_{\text{ss}}^{\text{II}})},
\end{equation}

The divergence of the bosonic approximation is more serious here than for $\hat{S}_y$, so more subtractions are needed. Through trial and error we found that it is convenient to express the numerator as
\begin{equationS}
    \mathrm{Tr}[\hat{\rho}_{\text{ss}}^{\text{II}}(\eta)\hat{S}_x^2]&=\mathrm{Tr}\bigg[\hat{\rho}_{\text{ss}}^{\text{II}}(\eta)\hat{S}_x^2-\hat{\rho}_{\text{ss}}^{\text{II}}(0)\bigg(\hat{S}_x^2+\frac{i\eta N^{1/3}}{ 32^{1/3}}\hat{S}_x\bigg)\bigg]\\[5pt]
    &+\mathrm{Tr}\bigg[\hat{\rho}_{\text{ss}}^{\text{II}}(0)\bigg(\hat{S}_x^2+\frac{i\eta N^{1/3}}{32^{1/3}}\hat{S}_x\bigg)\bigg].
\end{equationS}
The first line gives a finite result within the boson approximation, while the second line (which diverges in the boson treatment) can be calculated using semiclassical analysis. We quote here the final result
\begin{equationS}\label{eqn:AppSx2Fin}
    \braket{\hat{S}_x^2}=\bigg(\frac{N}{4}\bigg)^{2/3}\Bigg[&\frac{\int_0^{\infty}\,dv\,e^{-v^6/6}(e^{-2\eta v^2}-1)v^2}{2\int_0^{\infty}\,dv\,e^{-v^6/6-2\eta v^2}}\\[5pt]
    &+\frac{N/3}{\sqrt{2\pi}\int_0^{\infty}\,dv\,e^{-v^6/6-2\eta v^2}}\Bigg].
\end{equationS}

When $\eta\leq -1$, use of the stationary phase approximation leads to Eq.~(\ref{eqn:CRFSxVariance}), but Eq.~(\ref{eqn:AppSx2Fin}) is valid down to $\eta=0$ ($\Upsilon=1$). We see this in Fig.~\ref{fig:AppUniformApproximations}(c), which compares it against exact numerical simulations.

\end{document}